\documentclass[11pt]{article}

\usepackage{amsmath,amssymb,amsfonts}
\usepackage{physics}
\usepackage{relsize}
\usepackage{braket}
\usepackage{threeparttable}
\usepackage{enumerate}
\usepackage{bbm}
\usepackage{comment}

\usepackage{cases}

\usepackage{tikz}
\usetikzlibrary{cd}
\usetikzlibrary{shapes.geometric}
\usetikzlibrary{positioning}

\usepackage{cite}
\usepackage{subfiles}

\usepackage[top=30truemm,bottom=40truemm,left=30truemm,right=30truemm]{geometry}

\newcommand{\BB}{{\mathcal{B}}}
\newcommand{\CC}{{\mathcal{C}}}
\newcommand{\DD}{{\mathcal{D}}}

\newcommand{\OO}{{\mathcal{O}}}
\newcommand{\TT}{{\mathcal{T}}}
\newcommand{\II}{{\mathcal{I}}}

\newcommand{\NN}{{\mathcal{N}}}

\newcommand{\bj}{{\bar \jmath}}

\def\half{{\frac{1}{2}}}

\makeatletter

\@addtoreset{equation}{section}

\makeatother
\def\etitle#1{\def\@etitle{#1}}
\def\department#1{\def\@department{#1}}
\def\majoring#1{\def\@majoring{#1}}

\DeclareFontShape{OT1}{cmr}{mx}{n}%
{<->cmr10}{}

\DeclareMathAlphabet{\titlemath}{OT1}{cmr}{mx}{n}

\allowdisplaybreaks

\begin{document}

	\begin{titlepage}

		\begin{flushright}
			OCU-PHYS 522 
		\end{flushright}

		\vskip.5cm
		\begin{center}
		{\huge\fontseries{mx}\selectfont
			The Chiral Algebra of Genus Two \\
			Class $\mathcal{S}$ Theory \par}

\vskip 1.1cm
		
		{\large Kazuki Kiyoshige $^{\diamondsuit,1}$
			and Takahiro Nishinaka$^{\heartsuit,2}$}
		
		\vskip 1cm
		
		{\it
			$^1$Department of Physics, Graduate School of Science, Osaka City University\\
			3-3-138 Sugimoto, Sumiyoshi, Osaka 558-8585, Japan \\[2mm]
			$^2$Department of Physical Sciences, College of Science and Engineering\\
			Ritsumeikan University, Shiga 525-8577, Japan
		}

		\end{center}

		\vskip1.5cm

		\begin{abstract}
\vskip.3cm

			We  construct the chiral algebra associated with the $A_{1}$-type class $\mathcal{S}$ theory 
		for the genus two Riemann surface without punctures. By solving the BRST cohomology problem corresponding to a marginal gauging
		 in four dimensions, we find a set of chiral algebra generators that form closed OPEs. Given the fact that they reproduce the spectrum
		 of chiral algebra operators up to large dimensions, we conjecture that they are the complete
		 set of generators.
Remarkably, their OPEs are invariant under an action of $SU(2)$ which is not associated with any conserved
		 one-form current in four dimensions. We find that this novel
		 $SU(2)$ strongly constrains the OPEs of non-scalar
		 Schur operators. 
For completeness, we also check the equivalence of Schur indices
		 computed in two S-dual
		 descriptions with a non-vanishing flavor fugacity turned on.

		\end{abstract}

\renewcommand{\thefootnote}{\fnsymbol{footnote}}
\footnotetext{
	$^{\diamondsuit}$kiyoshig@sci.osaka-cu.ac.jp, 
	$^{\heartsuit}$nishinak@fc.ritsumei.ac.jp}
\renewcommand{\thefootnote}{\arabic{footnote}}

	\end{titlepage}

	\tableofcontents

			\section{Introduction}
\label{sec:intro}
			
			One of the most interesting recent developments in the study of four-dimensional $\NN=2$ superconformal field theories (SCFTs) is the new duality proposed by \cite{Beem:2013sza},
			\begin{align}
			\chi: \text{4d} ~ \NN=2\text{ SCFT } \to \text{2d} ~ \text{chiral algebra}\,,
			\label{eq:map4d2d}
			\end{align}
			which associates a 2d chiral algebra to every $\mathcal{N}=2$ SCFT in
			four dimensions.\footnote{In mathematical
			literature, chiral algebras are also called vertex operator algebras (VOAs).}
		This map can be applied to arbitrary four-dimensional $\NN=2$ SCFT, regardless of the existence of a
		weakly-coupled Lagrangian description.
		The associated chiral algebra constructed in this way 
		captures the OPEs of protected BPS operators called 
		``Schur operators'', which are 
		operators contributing to the Schur limit of the superconformal index
		\cite{Gadde:2011ik,Gadde:2011uv}. 
		While the scalar Schur operators are known as Higgs branch operators and well-studied, the nature of their 
		non-scalar cousins is not generally understood.

			 Associated chiral algebras have revealed various general properties of 4d $\mathcal{N}=2$ SCFTs.
			 Indeed, in \cite{Beem:2013sza,Lemos:2015orc,Liendo:2015ofa,Ramirez:2016lyk,Beem:2018duj}, many new 
			 unitarity bounds were found by using this 2d/4d correspondence. 
			 It was also shown in \cite{Beem:2017ooy} that
			 the  Higgs branch of the 4d theory is reconstructed as the
			 associated variety \cite{Arakawa1} of the corresponding chiral algebra, which 
			 was further studied in 
			 \cite{Arakawa:2016hkg,Arakawa:2017fdq,Bonetti:2018fqz, Arakawa:2018egx,Beem:2019tfp}.
			 Since the map \eqref{eq:map4d2d} preserves the non-perturbative structure 
			 of OPEs among the Schur operators, the associated chiral algebra is also a powerful tool to analyze strongly 
			 coupled SCFTs, such as Argyres-Douglas type theories
			 \cite{Argyres:1995jj,Argyres:1995xn,
			 Eguchi:1996vu, Xie:2012hs} and $T_N$ theories \cite{Gaiotto:2009we}.
			See	\cite{Beem:2014rza, Lemos:2014lua, Beem:2014zpa, 
Buican:2015ina,Cordova:2015nma,Buican:2015tda,Cecotti:2015lab, Beem:2016cbd, Nishinaka:2016hbw, 
Buican:2016arp, Xie:2016evu,Cordova:2016uwk, Lemos:2016xke,Beem:2016wfs,Bonetti:2016nma,
Song:2016yfd, Creutzig:2017qyf, Fredrickson:2017yka, Cordova:2017mhb, Song:2017oew,Buican:2017fiq,
 Neitzke:2017cxz, Pan:2017zie, Fluder:2017oxm,
Choi:2017nur,Buican:2017rya, Wang:2018gvb, 
Nishinaka:2018zwq,Agarwal:2018zqi,Kiyoshige:2018wol,
Buican:2019huq,	Xie:2019yds,
Pan:2019bor,Oh:2019bgz,Jeong:2019pzg,Dedushenko:2019yiw,
Xie:2019zlb,Beem:2019snk,Xie:2019vzr,
 Wang:2020oxs} for recent works in this and other directions.

			After these works on the associated chiral algebras, one important
			thing that is still to be understood is the chiral algebra nature of
			non-scalar Schur operators. In particular, chiral algebra generators
			corresponding to non-scalar Schur operators in the semi-short multiplet,
			$\widehat{\mathcal{C}}_{R(j,\bj)}$, are not well-understood, except for the stress tensor multiplet
			$\widehat{\mathcal{C}}_{0(0,0)}$.\footnote{Here, we use the convention
				of \cite{Dolan:2002zh}.} Most of
			the earlier works focused on the case in which the chiral algebra is
			generated by the stress tensor and those corresponding to
			Higgs branch operators (i.e., scalar Schur operators), possibly with
			their partners created by extra supersymmetries.\footnote{In some cases,
				the 2d stress tensor is a composite
				operator and therefore not an independent generator.} In such cases, the OPEs of the associated chiral
			algebras are often fixed by the 4d Higgs branch
			chiral ring and 2d Jacobi identities. However, there are SCFTs
			whose chiral algebra
			contains a generator corresponding to generic
			$\widehat{C}_{R(j,\bj)}$.\footnote{See
				\cite{Lemos:2014lua, Buican:2016arp} for examples of such chiral algebras.} 
			In this case, 2d OPEs are not simply reconstructed from the 4d Higgs branch.
			
			One simple such SCFT is the $A_1$-type class $\mathcal{S}$
			theory for a genus two Riemann surface without punctures.
			Here, $A_1$-type class $\mathcal{S}$ theories are defined as the IR limit of the six-dimensional $\NN=(2,0)\, A_1$ 
			SCFT 
			compactified on a punctured Riemann surface, $C_{g,s} $, where $g$ is genus and $s$ is the number of 
			punctures.\footnote{When the six-dimensional theory is
				$\NN=(2,0)\,\mathfrak{g}$ SCFT for general $\mathfrak{g}\in
				\{A_n,D_n,E_n\}$, the class $\mathcal{S}$ theory is also characterized
				by embeddings $\mathfrak{su}(2)\hookrightarrow \mathfrak{g}$ at each
				puncture on $\mathcal{C}_{g,s}$. In the case of $\mathfrak{g}=A_1$,
				there is only one such embedding in addition to the trivial one.} 
			We denote this theory by $\mathcal{T}_{\mathcal{C}_{g,s}}$, especially focusing on
			$\mathcal{T}_{C_{2,0}}$. 
			The flavor $U(1)_f$ symmetry of  $\mathcal{T}_{\mathcal{C}_{2,0}}$ is emergent and not associated with any puncture on the Riemann surface.  
			This is an example of flavor enhancement of the class $\mathcal{S}$ theories. 
			While $\mathcal{T}_{C_{2,0}}$ has a Lagrangian description, its associated chiral algebra has not been
			identified.\footnote{Note that, while chiral algebras of class $\mathcal{S}$ at
				genus zero are well-studied in \cite{Arakawa:2018egx} even for a general
				Lie algebra of the 6d (2,0) theory, its generalization to higher genus
				cases is still to be understood. See \cite{Yanagida:2020kim} for the
				derived extension of \cite{Arakawa:2018egx}, which will be helpful for
				understanding the chiral algebra interpretation of gluings that increase
				the genus of the Riemann surface.} Moreover, it was proven in \cite{Beem:2013sza} that its
			chiral algebra must contain generators corresponding to non-scalar Schur operators in  
			$\widehat{\mathcal{C}}_{1(0,0)}$. These operators are neither related
			to the stress tensor multiplet nor to Higgs branch operators, and still
			to be understood. For example,
			it is not known whether these generators are Virasoro primary
			operators.

			In this paper, we identify the chiral algebra $\chi[\mathcal{T}_{C_{2,0}}]$ associated
			with $\mathcal{T}_{C_{2,0}}$. Our strategy is to start
			with a weakly-coupled description of the theory and apply to it the
			2d interpretation of marginal gaugings proposed in
			\cite{Beem:2013sza}. This 2d interpretation involves a BRST reduction
			associated with a 4d marginal gauging. We evaluate this BRST reduction and find a set
			of chiral algebra generators equipped with closed OPEs. These generators
			are of dimension less than or equal to three, and reproduce the correct
			operator spectrum by normal-ordered products at least up to dimension six.
			Given this fact and their closed OPEs, we conjecture that they are the complete set of generators. 
			As a further consistency check, we compute the character
			of $\chi[\mathcal{T}_{C_{2,0}}]$ based on our conjecture, and find the result in perfect agreement with the Schur index of $\TT_{C_{2,0}}$ up to $\OO(q^9)$.
			The 4d superconformal multiplets corresponding to these generators are also identified. 
			We particularly show that the generators corresponding to
			$\widehat{\mathcal{C}}_{1(0,0)}$ are all Virasoro primaries.

			 One remarkable consequence of our OPEs is that there exists an unexpected $SU(2)$ acting on 
			 $\chi[\mathcal{T}_{C_{2,0}}]$ as an
			 automorphism. Since
			 $\chi[\mathcal{T}_{C_{2,0}}]$ has no 
			 $\widehat{\mathfrak{su}(2)}$ current, this $SU(2)$ symmetry is not associated with any 2d conserved 
			 current. 
			 Moreover, one can show that there is no 4d conserved one-form current corresponding to this 
			 $SU(2)$.\footnote{ By ``conserved one-form current'', we mean $j =j_\mu dx^\mu$ such that
$\partial_\mu j^\mu =0$, as in \cite{Gaiotto:2014kfa}. The corresponding Noether charge is given by $\int *j$.
}
			 Therefore, this either corresponds to a 4d symmetry without conserved current, or is an accidental symmetry 
			 in two dimensions.
			 We find that the action of this $SU(2)$ is trivial on the space of 2d operators corresponding to scalar Schur
			 operators. Therefore, it is a symmetry characterizing the OPEs
			 of non-scalar Schur operators. Indeed, the invariance
			 under this $SU(2)$ forbids various otherwise-possible terms in the OPEs of
			 these operators.\footnote{Such a symmetry is also
			 	present in the chiral algebra studied in \cite{Buican:2016arp}.}
			 It would be an
			 interesting open problem to see if this new symmetry corresponds to
			 a 4d global symmetry without conserved current.

			The organization of this paper is as follows. 
			In section \ref{sec:review}, we will briefly review the 4d/2d duality and the BRST interpretation of 4d 
			marginal gaugings, following \cite{Beem:2013sza}.
			In section \ref{sec:gesnu2}, we will collect aspects of the genus two theory $\TT_{C_{2,0}}$, and describe our strategy of identifying the corresponding chiral algebra $\chi[\TT_{C_{2,0}}]$. 
			Sections \ref{sec:ope} and \ref{sec:auto} include our main results. 
			In section \ref{sec:ope},
			we identify the chiral algebra generators and their OPEs, and in section \ref{sec:auto} we 
			discuss automorphisms of $\chi[\mathcal{T}_{C_{2,0}}]$ including a new $SU(2)$ symmetry that is not 
			associated with any 4d conserved one-form current.
			The final section \ref{sec:conclusion} is devoted to the conclusion and discussions about future works.
			While the genus two theory has two independent weak coupling descriptions, we focus on one of them. 
			The S-dual equivalence between them is checked via the superconformal index with its flavor fugacity turned 
			off 
			\cite{Gadde:2011ik}.
			In appendix \ref{app:index}, we provide its extension to the case of non-vanishing flavor fugacity.
			In appendix \ref{app:null}, we list the null operators in $\chi[\TT_{C_{2,0}}]$ up to dimension six.

			\section{Brief review of the chiral algebra conjecture } \label{sec:review}
			In this section, we will review the duality between four-dimensional $\NN=2$ SCFTs and 2d
			chiral algebras,
			following \cite{Beem:2013sza} (see also \cite{Lemos:2020pqv} for a lecture note). 
			Readers familiar with this 2d/4d correspondence can skip this
			section. We follow the convention of \cite{Beem:2013sza} unless
			otherwise stated.

			\subsection{General properties of associated chiral algebras}
\label{subsec:general}

The map \eqref{eq:map4d2d} from 4d $\mathcal{N}=2$ SCFTs to chiral
algebras was defined in \cite{Beem:2013sza} by considering cohomology
classes of local operators with respect to a particular linear
combination of Poincar\'e and conformal supercharges. At the origin, each cohomology
classes is represented by a local operator
annihilated by
$\mathcal{Q}_-^1,\,\tilde{\mathcal{Q}}_{2\dot-},\,\mathcal{S}_1^-$ and
$\tilde{\mathcal{S}}^{2\dot-}$, which is called a Schur operator. The 4d
unitarity implies that the quantum numbers of Schur operators satisfy
\begin{align}
 				E=j+\bj+2R~,\qquad 
				r=\bj-j~,
\end{align}
where $E$ is the conformal dimension, $(j,\bj)$ is the Lorentz spin, $R$
is the $SU(2)_R$ charge, and $r$ is the $U(1)_r$ charge.

Schur operators are classified by superconformal multiplets containing
			them. We call an $\mathcal{N}=2$ superconformal
			multiplet containing a Schur operator, a ``Schur
			multiplet''. In the language of \cite{Dolan:2002zh},
			the only Schur multiplets are
			$\hat{\BB}_{R},
			{\DD}_{R(0,\bj)},\bar{\DD}_{R(j,0)}$ and
			$\hat{\CC}_{R(j,\bj)}$, where $R,j,\bj$ are
			quantum numbers of the bottom component in each multiplet.
			The Lorentz spin and $U(1)_r$ charge of a Schur operator in
			these multiplets are shown in Table
			\ref{list:shcurops}, together with the
			holomorphic dimensions of the corresponding 2d
			operators. Note that scalar Schur
			operators are only in $\hat{\mathcal{B}}_{R}$.
			\begin{table}[t] \centering
				\begin{tabular}{|c|c|c|c|}
					\hline 
					Multiplet &$h$ & spin & $U(1)_r$ \\ 
					\hline 
					$\hat{\BB}_{R}$ &$R$& $(0,0)$ & 0\\
					\hline 
					${\DD}_{R(0,\bj)}$ &$R+\bj+1$&
					 $(0,\bj+ \half )$ & $\bj + \half $ \\
					\hline 
					$\bar{\DD}_{R(j,0)}$  & $R+j+1$
				     & $(j+ \half ,0)$ & $-j - \half $ \\
					\hline 
					$\hat{\CC}_{R(j,\bj)}$
				 &$R+j+\bj+2$& $(j+ \half ,\bj+ \half )$ & $\bj-j$ \\
					\hline 
				\end{tabular} 
				\caption{The list of Schur multiplets in \cite{Dolan:2002zh} notation. 
					$h$ and $r$ are a 2d holomorphic dimension and $U(1)_{r}$ charge of Schur operators, respectively.}
				\label{list:shcurops}
			\end{table}

The OPEs of the chiral algebra $\chi[\mathcal{T}]$ associated with a 4d
			$\mathcal{N}=2$ SCFT $\mathcal{T}$ is determined by OPEs of Schur operators in $\mathcal{T}$. 
			Since the map from 4d OPEs to 2d ones involves twisting the translation with $SU(2)_R$ rotation, the $SU(2)_R$ symmetry of $\mathcal{T}$ is not preserved by the OPEs of $\chi[\mathcal{T}]$. 
			On the other hand, the flavor and $U(1)_r$ symmetries of $\mathcal{T}$ give rise to a conserved global symmetry of $\chi[\mathcal{T}]$.

The associated chiral algebra $\chi[\mathcal{T}]$ contains special
sub-algebras uniquely fixed by the 4d symmetry of $\mathcal{T}$.
In particular, the Virasoro sub-algebra of $\chi[\mathcal{T}]$ arises from the self-OPEs
of the $SU(2)_R$ current in the stress tensor multiplet, $\hat{\mathcal{C}}_{0(0,0)}$. 
It is for this reason that the 2d Virasoro central charge, $c_{2d}$, is fixed by a 4d Weyl anomaly
coefficient, $c_{4d}$, as\footnote{Here, we normalized $ c_{4d}$ such that a free hypermultiplet has $ 
c_{4d}=\frac{1}{12}$.}
\begin{align}
 c_{2d} = -12 c_{4d}~.
\end{align}
When $\mathcal{T}$ has a flavor symmetry $G$, the flavor current
multiplet $\hat{\mathcal{B}}_1$ gives rise to the following affine Kac-Moody $G$
sub-algebra in $\chi[\mathcal{T}]$:
\begin{align}
				J^{A}(z)J^{B}(0)\sim\frac{k_{2d}\delta^{AB}}{z^2}+\frac{i f^{AB}{}_{C}J^{C}(0)}{z}\,,  
\end{align}
where $f^{AB}{}_C$ is the structure constant of $G$, and
$A,B,C=1,\cdots, \text{dim}\,G$. The level, $k_{2d}$, of the affine
Kac-Moody algebra is fixed by the 4d flavor central charge, $k_{4d}$, as\footnote{We use the convention that the 
$k_{4d} = 2$ for a
hypermultiplet in the fundamental representation of $G$.}
\begin{align}
 k_{2d} = - \half k_{4d}\,.
\end{align}
This implies that every 4d global symmetry that commutes with
the $\mathcal{N}=2$ superconformal symmetry gives rise to a
corresponding affine current in $\chi[\mathcal{T}]$ unless the 4d
symmetry has no conserved current.

The OPEs of the other 2d operators are more non-trivial, reflecting the
4d OPEs of Schur operators that are not fixed by the global
symmetry of $\mathcal{T}$. Among these operators, those corresponding to
$\hat{\mathcal{B}}_R$ 
are relatively well-understood, since the corresponding Schur operators
are Higgs branch operators whose vacuum expectation values parameterize
the Higgs branch of $\mathcal{T}$. Indeed, the OPEs of 2d operators
corresponding to $\hat{\mathcal{B}}_R$ are often reconstructed from the
Higgs branch chiral ring of $\mathcal{T}$. The Hall-Littlewood (HL)
chiral ring is an extension of the Higgs branch chiral ring by Schur
operators in $\bar{\mathcal{D}}_{R(j,0)}$.\footnote{These operators are
those contributing to the HL limit of the superconformal index \cite{Gadde:2011uv}.} Similarly, HL anti-chiral
ring is defined for $\mathcal{D}_{R(0,\bj)}$ and
$\hat{\mathcal{B}}_R$. A common feature of these multiplets,
$\bar{\mathcal{D}}_{R(j,0)},\,\mathcal{D}_{R(0,\bj)}$ and
$\hat{\mathcal{B}}_R$, is that the corresponding 2d operators are
guaranteed to be Virasoro primary operators.

			When the 4d theory $\TT$ has a Lagrangian
			description, Schur operators in the $
			{\DD}_{R(0,\bj)}$ and $\bar{\DD}_{R(j,0)}$ multiplets
			are gauge invariant composite operators involving at least one gaugino in a vector multiplet.
			Therefore, for Lagrangian theories, 2d operators corresponding to
			$\bar{\mathcal{D}}_{R(j,0)}$ or $\mathcal{D}_{R(0,\bj)}$ are always nilpotent with respect to
			the normal ordered product.
			The simplest example of $ {\DD}_{R(0,\bj)}\,, \bar{\DD}_{R(j,0)}$ type
			multiplets is an $\NN=2$ free vector multiplet,
			${\DD}_{0(0,0)}\oplus \bar{\DD}_{0(0,0)} $, whose associated chiral algebra is the small $(b,c)$-ghost 
			system.\footnote{The ``small'' means there are no $c_{0}$ mode in the spectrum, and on the other
			hands, full $(b,c)$-ghost system contain
			$c_{0}$.} Its OPE is expressed in terms of $\lambda:=b, \tilde{\lambda}:=\partial c$ as
			\begin{align}
				\lambda(z)	 \tilde{\lambda} (0)\sim \frac{1}{z^2}\,.
			\end{align}
			It is also known that, in $A_{1}$ type class $\mathcal{S}$
			theories, $\bar{\DD}_{R(j,0)}$ and $\bar{\DD}_{R(0,\bj)}$ exist only if the genus of the Riemann surface is 
			non-zero. 
			
			The last type of Schur multiplet, $
			\hat{\CC}_{R(j,\bj)}$, is not well-understood,
			except for the stress tensor multiplet
			$\hat{\mathcal{C}}_{0(0,0)}$.
			As mentioned above, the Schur operator in
			$\hat{\mathcal{C}}_{0(0,0)}$ is the highest weight component of
			the $SU(2)_{R}$ conserved current, and
			corresponds to the 2d stress tensor in the associated chiral algebra. 
			The chiral algebra nature of the other $\hat{\mathcal{C}}$-type multiplets are, however, not very clear. 
			Indeed, unlike the other Schur multiplets, a chiral algebra generator corresponding to 
			$\hat{\mathcal{C}}_{R(j,\bj)}$ is 
			not guaranteed to be a Virasoro primary operator.

			\subsection{BRST reduction corresponding to gauging}\label{subsec:BRST}

			Let $\TT$ be a 4d $\NN=2$ SCFT with a flavor symmetry $G$ whose flavor
			central charge is $k_{4d}=4h^{\vee}$, where $h^{\vee} $ is the dual Coxeter number of $G$.
			 In this case, we can construct a new $\NN=2$ SCFT, $\TT_{G}$, by marginally gauging the flavor $G$ 
			 symmetry of $\mathcal{T}$. 
			 Indeed, $k_{4d} = 4h^\vee$ precisely coincides with the condition for a vanishing $\beta$-function.

			On the associated chiral algebra side, this	procedure corresponds to considering a BRST cohomology of the tensor product of
			$\chi[\mathcal{T}]$ and the small $(b,c)$-ghost
			system in the adjoint representation of $G$.
			To describe this, let us consider the following BRST charge
			\begin{align}
			 Q_{\text{BRST}}:=\oint \frac{dz}{2\pi i} J_{\text{BRST}}\,, \quad 
			 J_{\text{BRST}}:= c_{A}\left(J^{A}+\half
			 J^{A}_{\text{gh}}\right)\,,
			\label{eq:BRST}
			\end{align}
			where $J^{A}$ for $A=1,\cdots, \text{dim}\, G$
			is the affine $G$ currents in $\chi[\TT]$, and 
			\begin{align}
			J^{A}_{\text{gh}}:=- i f^{ABC}b^{B}c^{C}
			\label{eq:Jgh}
			\end{align}
			is the affine $G$ currents in the $(b,c)$-ghost sector. 
			Note that $k_{4d} = 4h^\vee$ implies the affine currents $J^A$ have level $k_{2d} = -2h^\vee$, which is 
			precisely the condition that $Q_\text{BRST}^2 = 0$. 
			When the gauge coupling is turned off, the chiral algebra of the $\mathcal{T}_G$ is given by
			\begin{align}
				\left\{\,|\psi\rangle \in \chi[\mathcal{T}] \otimes (b^A, \partial c_A)\;|\;
				J^A_{\text{tot},0}|\psi\rangle = 0\, \right\}~,
			\end{align}
			where $J^A_{\text{tot},0}$ is the zero-mode of $J^A_\text{tot} :=J^A +
			J^A_\text{gh}$, and $(b^A,\partial c_A)$ stands for the small
			$(b,c)$-ghost system associated with $G$. The constraint
			$J^A_{\text{tot},0}|\psi\rangle=0$ corresponds to the Gauss law constraint that is present even for the zero gauge coupling. 
			When the gauge coupling is turned on, some Schur operators are lifted to non-Schur operators, giving rise to a reduced chiral algebra. 
			This reduced algebra was conjectured in \cite{Beem:2013sza} to be given by the $Q_\text{BRST}$-cohomology 
			\begin{align}
				\chi[\TT_{G}] =
			 H_{Q_{\text{BRST}}}\Big( \big\{\ket{\psi} \in
			 \chi[\TT] \otimes (b^A,\partial c^A) \;\big|\;
			 J^{A}_{\text{tot}\,0}\ket{\psi} = 0  \big\}
			 \Big)\,.
			\label{eq:BRST2}
			\end{align}
		As shown in \cite{Beem:2013sza}, the associated chiral algebra is
		independent of marginal couplings of the four-dimensional
		theory. Therefore \eqref{eq:BRST2} is identified as the chiral algebra
		of $\mathcal{T}_G$ at any generic value of the gauge coupling.

In Sec.~\ref{sec:ope}, we will use the above prescription to identify the chiral algebra $\chi[\TT_{C_{2,0}}]$ associated with genus two class $\mathcal{S}$ theory.

			\section{Genus two theory }\label{sec:gesnu2}

			 The main purpose of this paper is to identify the
			 chiral algebra of the genus two class $\mathcal{S}$
			 theory, $\mathcal{T}_{C_{2,0}}$,
			 obtained by compactifying the six-dimensional 
			 $\NN=(2,0)\,, A_1$ theory on a genus two Riemann surface $C_{2,0}$. 
			 Here we collect known facts about this theory,
			 mainly following Sec.~5.4 of \cite{Beem:2013sza}, so that
			 we can use them in identifying its chiral
			 algebra in the next section.
\subsection{Weak coupling descriptions}

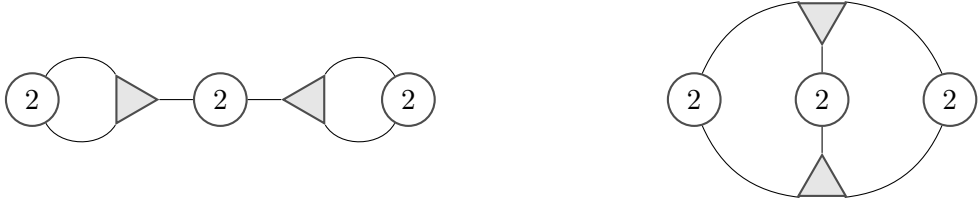
\begin{figure}[t]
				\begin{center}
					\begin{tikzpicture}
					[place/.style={circle, draw=black!70, thick,inner
						sep=0pt,minimum
						size=7mm},transition2/.style={rectangle,draw=black!50,fill=red!20,thick,innecr
						sep=0pt,minimum
						size=8mm},transition3/.style={rectangle,draw=black!70,thick,inner
						sep=0pt,minimum size=8mm}, triangle1/.style = {fill=gray!20, draw=black!70, thick, regular polygon, regular
						polygon sides=3, rotate=90}, triangle2/.style = {fill=gray!20, draw=black!70, thick, regular polygon, regular
						polygon sides=3, rotate=-90},triangle3/.style = {fill=gray!20, draw=black!70, thick, regular polygon, regular
						polygon sides=3, rotate=0},triangle4/.style = {fill=gray!20, draw=black!70, thick, regular polygon, regular
						polygon sides=3, rotate=180},auto]
					
					\node[place] (1) at (0,0) {\;$2$\;};
					
					\node[triangle1] (2) at (1.2,0) {} edge (1);
					\node[place] (3) at (2.5,0) {\;$2$\;} edge[bend right = 60] (2);
					\path[-] (3) edge[bend left = 60] (2);
					
					\node[triangle2] (4) at (-1.2,0) {} edge (1);
					\node[place] (5) at (-2.5,0) {\;$2$\;} edge[bend right=60] (4);
					\path[-] (5) edge[bend left=60] (4);

					\node[place] (a) at (8,0) {\;$2$\;};

					\node[place] (d) at (6.3,0) {\;$2$\;} ;
					\draw[-] (d) edge[bend left] (7.7,1.3);
					\draw[-] (d) edge[bend right] (7.7,-1.3);
					
					\node[place] (e) at (9.7,0) {\;$2$\;};
					\draw[-] (e) edge[bend left] (8.3,-1.3);
					\draw[-] (e) edge[bend right] (8.3,1.3);
					
					\node[triangle3] (b) at (8,-1.1) {} edge (a);
					\node[triangle4] (c) at (8,1.1) {} edge (a);

					\end{tikzpicture}
					\caption{Two weak-coupling descriptions of the theory $\TT_{\CC_{2,0}}$,
						which are S-dual to each other. Each circle node stands for an $SU(2)$
						gauge group, while each triangle node stands for the $T_{2}$ theory,
						i.e., the theory of four free hypermultiplets. Each corner
						of the triangles stands for an $SU(2)$ flavor subgroup of $T_{2}$.
						When a corner of a triangle is connected to a circle, the
						corresponding $SU(2)$ flavor subgroup of $T_{2}$ is coupled to
						the gauge group associated with the circle.
					}
					\label{fig:quiver1}
				\end{center}
			\end{figure}

			The $\TT_{\CC_{2,0}}$ theory has two different weak coupling descriptions, as shown in quiver gauge theories 
			in Fig.~\ref{fig:quiver1}, 
			corresponding to two different pants decompositions of $C_{2,0}$. 
			These two descriptions are expected to be S-dual to each other.
			In Fig.~\ref{fig:quiver1}, each circle node stands for an $SU(2)$ gauge group, while each triangle node stands 
			for the $T_{2}$ theory,	i.e., the theory of four free hypermultiplets. 
			The $T_{2}$ theory has $Sp(4)$ flavor symmetry, whose $SU(2)^3$ subgroup is manifest in 
			Fig.~\ref{fig:quiver1}; 
			each corner of the triangles stands for one $SU(2)$ flavor subgroup.
			When a corner of a triangle attaches to a circle, the corresponding $SU(2)$ flavor subgroup is gauged by the gauge group associated with the circle.

			The above $SU(2)^3$ flavor subgroup of the $T_{2}$
			theory can be explicitly seen as follows. In the
			$\mathcal{N}=1$ language, the $T_2$ theory is
			composed of eight chiral multiplets, which we
			denote by $Q_{ab}$ and $\widetilde{Q}^{ab}$ for
			$a,b=1,2$. There is an obvious $\mathcal{N}=2$
			flavor $SU(2)^2$ symmetry under which $Q_{ab}$
			and $\tilde{Q}^{ab}$ transform respectively as ${\bf 2} \otimes {\bf
			2}$ and $\overline{\bf 2}\otimes \overline{\bf
			2}$.\footnote{The indices, $a$ and $b$, express
			components of the fundamental or
			anti-fundamental representation of $SU(2)$ here.}
In addition, there exists an extra flavor $SU(2)$ symmetry under which $(Q_{1bc},Q_{2bc})$ defined by
			\begin{align}
			Q_{1bc} := Q_{bc}~,\qquad Q_{2bc} := \epsilon_{bd}\epsilon_{ce}\widetilde{Q}^{de}~,
			\end{align}
transforms as a fundamental representation, where $\epsilon_{ab}$ is the
anti-symmetric tensor such that $\epsilon^{12}=-\epsilon_{12} = 1$.
Therefore, $Q_{abc}$ can be regarded as a ``half-hypermultiplet'' in the trifundamental
			representation, ${\bf 2}\otimes {\bf 2}\otimes {\bf 2}$, of $SU(2)^3$
			\cite{Gaiotto:2009we}. 

As mentioned in section \ref{sec:intro}, the genus two theory
$\mathcal{T}_{C_{2,0}}$ has an accidental flavor $U(1)_f$ symmetry that is
not visible in its class $\mathcal{S}$ construction (see \cite{Maruyoshi:2009uk} for similar accidental 
enhancements of flavor symmetries of $A_1$-type class $\mathcal{S}$ theories). 
As a result, the most general 
expression for its superconformal index contains a fugacity for this flavor $U(1)_f$ symmetry. 
The explicit action of this $U(1)_f$ will be shown in sub-section \ref{subsec:Higgs-branch}.

The S-duality equivalence between the two quiver descriptions in Fig.~\ref{fig:quiver1}
has been checked in terms of the Schur limit of their superconformal indices with the flavor fugacity
turned off \cite{Gadde:2011ik}. The extension of their proof to the case
with a non-vanishing flavor fugacity is discussed in Appendix
\ref{app:index} of this paper. Given these results, we assume the
S-duality equivalence between the two quiver descriptions.
 In the rest of this paper, we mainly
focus on the left description in Fig.~\ref{fig:quiver1}, which we call
the ``dumbbell quiver''.

\subsection{Higgs branch and flavor symmetry}
\label{subsec:Higgs-branch}
		
In this sub-section, we describe the $U(1)_f$ flavor symmetry and a chiral
ring relation of the genus two theory
$\mathcal{T}_{C_{2,0}}$. We focus on the dumbbell quiver
description of the theory (i.e., the left quiver in Fig.~\ref{fig:quiver1}), and denote by $Q_{abc}$ and $S_{def}$ two
trifundamental half-hypermultiplets of $SU(2)^3$. 
		Let $SU(2)_1,\,SU(2)_2$ and $SU(2)_3$ be the gauge groups corresponding to the left, middle and
		right circles in the quiver, respectively. 
Without loss of generality, we say $SU(2)_1$ is diagonally gauging 
		the $SU(2)\times SU(2)$ symmetry corresponding to $b$
		and $c$ of $Q_{abc}$, and $SU(2)_3$ is diagonally
		gauging the
		$SU(2)\times SU(2)$ symmetry corresponding to $b$ and $c$ of
		$S_{abc}$.  The remaining gauge group $SU(2)_2$ is
		diagonally gauging $SU(2)$ corresponding
		to $a$ of $Q_{abc}$ and $S_{abc}$. 
		Then we see that, under $SU(2)_1$, the left
		half-hypermultiplet $Q_{abc}$ is decomposed into ${\bf
		1} \oplus {\bf 3}$ as \footnote{ The $\sigma$ matrices and antisymmetric tensor, $\epsilon $, are in the same convention with \cite{Wess:1992cp} and 
			$(\sigma^{A})_{ab}:= \epsilon_{ac}(\sigma^{A})^{c}{}_{b}, ~
			(\sigma^{A})^{ab}:=\epsilon^{bc} (\sigma^{A})^{a}{}_{c}$. 
			$(\sigma^{A})_{ab}$ are symmetric tensor. }
		\begin{align}
		Q_{a}:=\frac{1}{\sqrt{2}}\epsilon^{bc}Q_{abc}\,, \qquad 
		Q^{A}_{a}:=\sigma^{Abc} Q_{abc} \,.
		\end{align}
		Similarly, $S_{abc}$ also decomposes into ${\bf 1}\oplus
		{\bf 3}$ under $SU(2)_3$ as
		\begin{align}
		S_{a}:=\frac{1}{\sqrt{2}}\epsilon^{bc}S_{abc}\,, \qquad 
		S^{A}_{a}:=\sigma^{Bbc} S_{abc} \,.
		\end{align}
Note that these are all in the fundamental representation of $SU(2)_2$.
It is also useful for us below to define 
\begin{align}
 \phi_a :=\frac{1}{\sqrt{2}}\left(Q_a + iS_a\right) ~, \qquad
 \bar{\phi}_a := \frac{1}{\sqrt{2}}(Q_a-iS_a)~.
\label{eq:phi}
\end{align}		
The superpotential of the dumbbell quiver is then written as
\begin{align}
 W_{C_{2,0}} &= 4i f_{AB}{}^CQ^A_{a}Q^B_{b}\Phi^{(1)}_C
 \epsilon^{ab} + 4if_{DE}{}^FS^D_dS^E_e\Phi^{(3)}_F\epsilon^{de}
\nonumber\\[1mm]
&\qquad +2(Q_a^AQ_{Ab} + S_a^DS_{Db} + \phi_a
 \bar{\phi}_b)\sigma^{Aab}\Phi^{(2)}_{A}~,
\label{eq:superpot}
\end{align}
where $\Phi^{(i)}_A$ is the chiral superfield arising from the vector multiplet
associated with $SU(2)_i$.

From the above superpotential, we see that there is a flavor $U(1)_f$
symmetry under which $\phi_a$ and $\bar{\phi}_a$ have charge $+1$ and
$-1$ while all the other fields are neutral. The generators of the
Higgs branch chiral ring can also be read off from the above
superpotential. Indeed, it is generated by the moment map of
$U(1)_f$\footnote{The flavor moment map is a Schur operator in the
flavor current multiplet $\widehat{\mathcal{B}}_1$.}
			 \begin{align}
			 M:=-\phi_{a}\bar{\phi}^{a}\,,
			 \end{align}
			and the following Schur operators in the $\widehat{\BB}_{2}$ multiplet:
			\begin{align}
					\OO_{1}:=2\phi_{a}{\phi}_{b}
					Q_{A}^aQ^{Ab}
					\,,\qquad
					\OO_{2}:=2\bar{\phi}_{a}\bar{\phi}_{b}
					Q_{A}^aQ^{Ab}~,
					\label{eq:O1O2}
			\end{align}
where $\bar\phi^a := \epsilon^{ab}\bar\phi_b$ and $Q^a_A := \epsilon^{ab}Q_{Ab}$.
These Higgs branch operators satisfy the chiral ring relation
\begin{align}
 \mathcal{O}_1 \mathcal{O}_2 = M^4~,
\label{eq:higgrel}
\end{align}
as seen from the superpotential \cite{Hanany:2010qu}.

Note  that, when taking the decoupling limit of the middle gauge group $SU(2)_{2}$, 
the second line of \eqref{eq:superpot} vanishes.
Then the theory splits into the following three sectors:
\begin{itemize}
 \item  $\mathcal{N}=4\; SU(2)$ super Yang-Mills (SYM) theory described
	by $(Q_a^A,\Phi_C^{(1)})$.
 \item $\mathcal{N}=4\; SU(2)$ SYM theory described by
       $(S_a^A,\Phi_C^{(3)})$.
 \item a fundamental hypermultiplet of $SU(2)$ described by
       $(\phi_a,\bar{\phi}_a)$.
\end{itemize}
This means that the genus two theory $\mathcal{T}_{C_{2,0}}$
can be regarded as a theory obtained by marginally gauging a diagonal $SU(2)_2$ global symmetry of the above 
three sectors.\footnote{Note that $\mathcal{N}=4$ SYM theory has an $\mathcal{N}=2$ flavor $SU(2)$ symmetry.} 
We will use this construction of $\mathcal{T}_{C_{2,0}}$ in identifying its associated chiral algebra in the next section.

 			\section{Chiral algebra of genus two theory }
			\label{sec:ope}

In this section, we construct the chiral algebra
$\chi[\mathcal{T}_{C_{2,0}}]$ associated with the genus two
theory $\mathcal{T}_{C_{2,0}}$. We regard $\mathcal{T}_{C_{2,0}}$ as a theory obtained by marginally
gauging the three sectors discussed at the end of the previous
section. Then $\chi[\mathcal{T}_{C_{2,0}}]$ is constructed via the
BRST reduction corresponding to this marginal gauging.

			 \subsection{Chiral algebras of the three sectors}

The chiral algebras of the three sectors discussed at the end of sub-section
\ref{subsec:Higgs-branch} are already identified. 

The chiral algebra of a fundamental hypermultiplet of $SU(2)$ is the
symplectic boson algebra generated by $\phi^{ai}$ for $a=1,2$ and $i=\pm$
such that
\begin{align}
\phi^{a\pm}(z) \phi^{b\mp}(0) \sim \frac{\epsilon^{ab}}{z}~.
\label{eq:SB}
\end{align}
Here, the 2d operators $\phi^{a+}$ and $\phi^{a-}$ correspond
respectively to the
4d Schur operators $\phi^{a}$ and $\bar{\phi}^a$ described in \eqref{eq:phi}.\footnote{The 2d
operators $q^{a,1}:=\frac{1}{\sqrt{2}}(\phi^{a+} + \phi^{a-})$ and
$q^{a,2}:= \frac{1}{\sqrt{2}i}(\phi^{a+}-\phi^{a-})$ correspond to
the 4d Schur operators $Q^a$ and $S^a$, respectively. Their 2d OPEs are
given by $q^{a,i}(z)q^{b,j}(0) \sim \epsilon^{ab}\delta^{ij}/z$.}
Recall that the 4d $U(1)_r$ symmetry is generally preserved by 2d OPEs. Since the 4d Schur operators $\phi^a$ and $\bar{\phi}^a$ are
neutral under $U(1)_r$, so are $\phi^{a\pm}$.
This chiral algebra has 
$\widehat{\mathfrak{sp}(4)}$ currents, which
contains as sub-algebras the 
$\widehat{\mathfrak{su}(2)}$ currents  
\begin{align}
 J^A_\text{matter} :=  - \half \sigma^A_{ab} \phi^{a+}\phi^{b-}~,
\label{eq:Jmatt}
\end{align}
 and the $\widehat{\mathfrak{u}(1)}$ current
\begin{align}
 J :=  \epsilon_{ab}\phi^{a+}\phi^{b-}~.
\end{align}
In the BRST reduction discussed below, $J^A_\text{matter}$
will be involved in the BRST current and therefore disappears from the
spectrum. On the other hand, $J$ will give rise to a non-trivial BRST
cohomology class corresponding to the flavor $U(1)_f$ symmetry of
$\mathcal{T}_{C_{2,0}}$.

Let us now turn to the other sectors, i.e.,
			 $\mathcal{N}=4\; SU(2)$ SYM theories. The
			 chiral algebra associated with the
			 $\mathcal{N}=4\; SU(2)$ SYM was conjectured in
			 \cite{Beem:2013sza} to be the small
			 $\mathcal{N}=4$ super Virasoro algebra at the
			 Virasoro central charge $c=-9$. This algebra is
			 generated by 
$\widehat{\mathfrak{su}(2)}$ currents
			 $J^A$ at level $-3/2$, and $\mathcal{N}=4$ supercurrents
			 $G^a,\,\bar{G}^a$. Their non-trivial OPEs are
			 given by
\begin{align}
 J^A(z) J^B(0) &\sim -\frac{3}{4}\frac{\delta^{AB}}{z^2} +
 \frac{if^{ABC}J_C}{z}~,
\\
J^A(z) G^a(0) &\sim - \half \frac{(\sigma^{A})^a{}_b G^b}{z}~,
\\
J^A(z)\bar{G}^a(0) &\sim - \half \frac{(\sigma^A)^a{}_b
 \bar{G}^b}{z}~,
\\
G^a(z)\bar{G}^b(0) &\sim -\frac{6\epsilon^{ab}}{z^3} +\frac{4(\sigma^{A})^{ab}J_{A}}{z^2}
						+\frac{2(\sigma^{A})^{ab}\partial J_{A}+2\epsilon^{ab} T_{\text{sug}}}{z}~,
\end{align}
where $f^{ABC} = \epsilon^{ABC}$ is the structure constant of
$SU(2)$, and $T_\text{sug} : = 2J_AJ_A$ is the Sugawara stress tensor. 
The $U(1)_r$ charges of $J^A,G^a$ and $\bar{G}^a$ are
respectively $0,\, \half $ and $- \half $.\footnote{The Schur operators
corresponding to $G^a$ and $\bar{G}^a$ are in
$\mathcal{D}_{ \half (0,0)}$ and
$\bar{\mathcal{D}}_{ \half (0,0)}$, respectively. Their $U(1)_r$
charges can be seen from Table
\ref{list:shcurops}. } 
Note that the above OPEs preserve this $U(1)_r$
symmetry as expected.

The small $\mathcal{N}=4$ Virasoro algebra at $c=-9$ has various null
operators, which we need to remove from the spectrum. 
Fortunately, there is a special free field realization of this algebra which makes all these null operators automatically vanishing \cite{Adamovic:2014lra}.
 As shown in \cite{Bonetti:2018fqz}, this free field realization is concisely expressed in terms of a $\beta\gamma b c$ ghost system. 
To describe it, let us change variables as 
$J^{\pm}:=-i\left( J^{1}\pm i J^{2} \right)\,,
			  J^{0}:=2 J^{3}\,,
			  G^+ :=-\frac{1+i}{2}G^2,\,
			 G^-:=
			 \frac{1-i}{2}G^1\,,
			 \bar{G}^+ :=-\frac{1+i}{2}\bar{G}^2$ and $\bar{G}^- := \frac{1-i}{2}\bar{G}^1$.
Then the free field realization is given by
			\begin{align*}
				J^{+}&=\beta\,,\quad
				J^{0}=bc +2\beta \gamma\,,\quad
				J^{-}=\beta \gamma\gamma+\gamma bc -\frac{3}{2}\partial \gamma~,\\
				G^{+}&=b^{i}\,,\quad
				G^{-}=b \gamma\,,\quad
				\bar{G}^{+}= c \partial
			 \beta +2 (\partial c) \beta\,,\quad
				\bar{G}^{-}=-b(\partial c) c+2\beta
			 \gamma \partial c+\partial \beta \gamma
			 c-\frac{3}{2}\partial^2c\,.
\label{eq:free}
			\end{align*}
Here the OPEs of the $\beta\gamma b c$ system are given by
\begin{align}
\beta(z)\gamma(0) \sim -\frac{1}{z}~,\qquad b(z)c(0) \sim \frac{1}{z}~.
\end{align}
In the next sub-section, we use this free field realization for each of the two $\mathcal{N}=4$ SYM sectors. 
Note that $b$ and $c$ that appear in the above free field realization have nothing to do with the $bc$  ghost system arising in the BRST reduction.

			 \subsection{BRST reduction}
\label{subsec:BRST2}

Let us now consider marginally gauging the three sectors to obtain the
genus two theory $\mathcal{T}_{C_{2,0}}$. As reviewed in
sub-section \ref{subsec:BRST}, on the chiral algebra side, this
corresponds to a BRST reduction of the tensor product of chiral
algebras associated with these three sectors. 

Recall that two of the three sectors are described by $\mathcal{N}=4\;
SU(2)$ SYM. For $k\in \{1,2\}$, let
$J^A_{(k)},G^a_{(k)}$ and $\bar{G}^a_{(k)}$ be the generators of the small
$\mathcal{N}=4$ super Virasoro algebra (at $c=-9$) associated with the
$k$-th $\mathcal{N}=4$ SYM sector. The remaining sector is the theory of
a fundamental hypermultiplet of $SU(2)$, whose chiral algebra is generated by the symplectic bosons $\phi^{a\pm}$ such that \eqref{eq:SB}.

In the tensor product of these three chiral algebras, 
\begin{align}
 J^A := J^A_{(1)} + J^{A}_{(2)} + J^A_\text{matter}
\end{align}
is the $\widehat{\mathfrak{su}(2)}$ current corresponding to the 4d $SU(2)$ symmetry that we are gauging, 
where $J^A_\text{matter}$ is given
by \eqref{eq:Jmatt}. With this $J^A$, the relevant BRST current is given by 
\begin{align}
 J_\text{BRST} = c_A\left(J^A +  \half J^A_\text{gh}\right)~,
\end{align}
where $J^A_\text{gh}$ is defined in \eqref{eq:Jgh}. We see that
$Q_\text{BRST} = \oint \frac{dz}{2\pi i}J_\text{BRST}$ is nilpotent, which reflects the fact that the corresponding $SU(2)$ gauging is
exactly marginal in four dimensions.

According to the gauging prescription reviewed in sub-section
\ref{subsec:BRST}, the chiral algebra
$\chi[\mathcal{T}_{C_{2,0}}]$ is identified as the following
BRST cohomology:
\begin{align}
 \chi[\mathcal{T}_{C_{2,0}}] = H_{Q_\text{BRST}} \Bigg( \left\{\;
 |\psi\rangle\in \left(\bigotimes_{k=1}^2
 (J^A_{(k)},G_{(k)}^a,\bar{G}^a_{(k)})\right)\otimes
 (\phi^{a\pm})\otimes (b^A,\partial c^A) \; \Bigg| \;
 J^A_{\text{tot},0} |\psi\rangle =0\;\right\} \Bigg)~,
\label{eq:coh}
\end{align}
where $(x,y,\cdots)$
stands for the chiral algebra generated by $x,y,\cdots$, and $J^A_{\text{tot}} := J^A + J^A_\text{gh}$. We evaluate
this cohomology explicitly, using the Mathematica
package \verb|OPEdefs| developed by \cite{Thielemans:1991uw, Thielemans:1994er}. While this
computation is rather involved, one can simplify it by using the free
field realization of $(J^A_{(k)}, G^a_{(k)},\bar{G}^a_{(k)})$ reviewed
in the previous sub-section.\footnote{One thing that makes this
computation involved is the presence of various null operators in
$(J^A_{(k)},G^a_{(k)},\bar{G}^a_{(k)})$ that we need to remove from the spectrum. Using the free field realization, 
all these null operators are
automatically zero, which simplifies the computation with 
Mathematica.} Evaluating the cohomology classes in \eqref{eq:coh} up to a high order of 
holomorphic dimensions of operators, we find that there exists a set, $\mathtt{S}$,
of operators in $\chi[\mathcal{T}_{C_{2,0}}]$ with the following properties.
\begin{itemize}
 \item The operators in $\mathtt{S}$ are of dimension less than or equal to three.

 \item Every
$Q_\text{BRST}$-cohomology class of dimension less than or equal to six
       is either an operator in $\mathtt{S}$, the derivative of an operator in
       $\mathtt{S}$, or
a normal ordered product of these operators.

 \item The operators in $\mathtt{S}$ form closed OPEs.
\end{itemize}
The list of operators in
$\mathtt{S}$ is shown in Table \ref{tb:generators} with their quantum
numbers and corresponding 4d superconformal multiplets.

\begin{table}[t] \centering
 
				\begin{tabular}{|c|c|c|c|c|}
					\hline 
					 2d generator &$h$ & $f$ & $r$ & 4d
					     multiplet  \\
					\hline\hline
					$J$& 1 & 0 & 0 & $\hat{\BB}_{1}$ \\ 
					\hline 
					$B^{+}$& 2 & 2 & 0 & $\hat{B}_{2}$ \\ 
					\hline 
					$B^{-}$& 2 & -2 & 0 & 
						 $\hat{B}_{2}$  \\ 
					\hline 
					$D^{+ I}$& 2 &
					     1 & $ \half $ & ${\DD}_{1(0,0)}$  \\ 
					\hline 
					$\bar{D}^{+}_{I}$& 2 & 1 &
					     $- \half $ & $\bar{\DD}_{1(0,0)}$ \\ 
					\hline 
					$D^{- I}$& 2 &
					     -1 & $ \half $ & ${\DD}_{1(0,0)}$ \\ 
					\hline 
				 $\bar{D}^{-}_{I}$& 2 & -1 &
					     $- \half $ & $\bar{\DD}_{1(0,0)}$  \\ 
					\hline 
					$T$& 2 & 0
					     & 0 & $\hat{\CC}_{0(0,0)}$ \\ 
					\hline 
					$X$& 3 & 0  & 1 & ${\DD}_{\frac{3}{2}(0,\half)}$\\ 
					\hline 
					$ \bar{X}$& 3 & 0 & $-1$ & $\bar{\DD}_{\frac{3}{2}(\half,0)}$ \\ 
					
					\hline 
					 $C_{A}$&
					 3 & 0 & 0 & $\hat{\CC}_{1(0,0)}$ \\ 
					\hline 
				\end{tabular} 
			 	\caption{The list of generators of genus two chiral algebra that we found. 
			 Here, $h$ is the holomorphic dimension of operators, $f$ is the global $U(1)$ 
			 charge associated with the $\widehat{\mathfrak{u}(1)}$ current $J$, and $r$  is the $U(1)_r$ charge. 
 			We conjecture that this is the complete set of generators. 
 			Note that the indices $I=\uparrow,\downarrow$ and $A=1,2,3$ have nothing to do with the global $U(1)$ 
 			symmetry. 
 			They play an important role in the study of automorphisms of $\chi[\mathcal{T}_{C_{2,0}}]$ 
 			in section \ref{subsec:newsym}. }
				 \label{tb:generators}
			\end{table}

Given these facts, we {\it conjecture} that the above $\mathtt{S}$ is the complete set of
generators of the chiral algebra
$\chi[\mathcal{T}_{C_{2,0}}]$. Similar
conjectures have been
made for various theories \cite{Beem:2013sza, Nishinaka:2016hbw, Buican:2017fiq}, leading to consistent
results. We will perform a further consistency check of our conjecture in
sub-section \ref{subsec:check}. In the next sub-section, we describe which $Q_\text{BRST}$-cohomology classes correspond
to the operators in $\mathtt{S}$.

Note that, as shown in \cite{Beem:2013sza}, the Macdonald index of
			$\mathcal{T}_{C_{2,0}}$ implies that
			there are {\it at least} those generators listed
			in Table \ref{tb:generators}. Although the index
			computation does not imply the absence of extra
			generators, we have shown here that there is no
			extra generators up to dimension six. Combined
			with the fact that they form closed OPEs, this
			is a strong evidence for the above conjecture.

			\subsection{List of generators of $\chi[\mathcal{T}_{C_{2,0}}]$}
\label{subsec:list}

We here describe the $Q_\text{BRST}$-cohomology classes corresponding to
the generators listed in Table \ref{tb:generators}, in terms of
their representatives.
Note that these cohomology classes are labeled by three quantum numbers $(h,r,f)$, 
where $h$ is the holomorphic dimension, $r$ is $U(1)_{r}$ charge, and $f$ is flavor $U(1)_{f}$ charge.

 First, at dimension one, the only generator is the 
$\widehat{\mathfrak{u}(1)}$ current
\begin{align}
 	J=\epsilon_{ab}\phi^{a+}\phi^{b- }\,.
\end{align}

At dimension two, there are three independent bosonic generators. The one with vanishing global $U(1)_f$ charge is the
stress tensor\footnote{Note that $T$ cannot be the Sugawara-type stress
tensor associated with $J$ since $c_{2d}=-26$.}
\begin{align}
T &=  \sum_{i=1}^2 T_{\text{sug}(i)}-  \half \epsilon_{ab}(\phi^{a+} \partial\phi^{b-} +
 \phi^{a-} \partial \phi^{b+}) - b^A\partial c_A ~,
\end{align}
where $T_{\text{sug}(i)} := J^A_{(i)}J_{(i)A}$.
Since the genus two theory
$\mathcal{T}_{C_{2,0}}$ has $c_{4d} = 13/6$, the Virasoro central charge is $c_{2d} = -26$.
The other two bosonic operators are written as
\begin{align}
 B^\pm = -2\left(J^A_{(1)}-J^A_{(2)}\right)\sigma_{Aab}\,\phi^{a\pm }\phi^{b\pm}~.
\end{align}
These operators have $U(1)_f$ charge $\pm 2$, and correspond to the 4d Higgs branch operators, $\mathcal{O}_1$ and
$\mathcal{O}_2$, shown in \eqref{eq:O1O2}.

At dimension two, there are also the following fermionic
generators:
\begin{align}
 D^{\pm \uparrow} &= -\epsilon_{ab}\phi^{a\pm}\left(iG_{(1)}^b \pm
 G_{(2)}^b \right)~,\qquad D^{\pm \downarrow} =
 -\epsilon_{ab}\phi^{a\pm}\left(iG_{(1)}^b \mp G_{(2)}^b\right)~,
\\
\bar{D}^\pm_{\uparrow} &=
 -\epsilon_{ab}\phi^{a\pm}\left(i\bar{G}_{(1)}^b \pm
 \bar{G}_{(2)}^b\right)~,\qquad \bar{D}^{\pm}_{\downarrow} =
 -\epsilon_{ab}\phi^{a \pm}\left(i\bar{G}_{(1)}^b \mp \bar{G}_{(2)}^b\right)~.
\end{align}
The $U(1)_f$ charge of $D^{\pm I}$ and $D^\pm_{I}$ are both $\pm 1$.
Here, one could instead take $\epsilon_{ab}\phi^{a\pm}G^{b}_{(i)}$ and
$\epsilon_{ab}\phi^{a\pm}\bar{G}^{b}_{(i)}$ as independent
operators, but we will see below that the above linear combinations
 make automorphisms of
$\chi[\mathcal{T}_{C_{2,0}}]$ more transparent. From their $U(1)_r$ charges, we see that the corresponding Schur
multiplets of $D^{\pm I}$ and $\bar{D}^\pm_I$ are as in Table \ref{tb:generators}.

At dimension three, there are five generators, which are all neutral
under the global $U(1)_f$ symmetry. We split them into two parts;
			\begin{align}
			X=i\epsilon_{ab}{G^{a}_{(1)}G^{b}_{(2)}}\,, \qquad 
			\bar{X}=- i \epsilon_{ab}{\bar{G}^{a}_{(1)}\bar{G}^{b}_{(2)}}\,, 
			\end{align}
and
	\begin{align}
			C_{1}&=4J^A_{(1)} J_{(1)A}\epsilon_{ab}\phi^{a+}\phi^{b-}+2J^A_{(1)}\sigma_{Aab}\left(\phi^{a+} \partial \phi^{b-}-\phi^{a-} \partial\phi^{b+}\right)
			-(1\leftrightarrow 2)\,,
			\label{def:C2}
\\
			C_{2}&=\epsilon_{ab}({G^{a}_{(1)}\bar{G}^{b}_{(2)}}+{G^{a}_{(2)}\bar{G}^{b}_{(1)}})\,,
\\
			C_{3}&= i\epsilon_{ab}({G^{a}_{(1)}\bar{G}^{b}_{(2)}}-{G^{a}_{(2)}\bar{G}^{b}_{(1)}})\,.
			\end{align}
From their $U(1)_r$ charges, we see that $X$ and
$\bar{X}$ correspond to
Schur operators in
$\mathcal{D}_{\frac{3}{2}(0, \half )}$ and 
$\bar{\mathcal{D}}_{\frac{3}{2}( \half ,0)}$, and similarly $C_A$ correspond
to $\widehat{\mathcal{C}}_{1(0,0)}$.

\subsection{OPEs of generators}
\label{subsec:OPEs}

Here, we describe the OPEs of generators listed in Table
\ref{tb:generators}. Note that these are uniquely fixed by the BRST
cohomology computation. First, the OPEs of $T$ and $J$ are given by
\begin{align}
 T(z) T(0) \sim -\frac{13}{z^4} + \frac{2T}{z^2} + \frac{\partial T}{z}~, \qquad 
J(z)J(0) \sim -\frac{2}{z^2}~,\qquad 
T(z)J(0) \sim \frac{J}{z^2} + \frac{\partial J}{z}~.
\end{align}
Recall here that the Virasoro central charge $c_{2d} =-26$ follows from
the 4d central charge $c_{4d}=13/6$.

Next, we find that all the other generators in Table \ref{tb:generators} are primary
operators in the sense of the Virasoro sub-algebra generated by $T$ and
the $\widehat{\mathfrak{u}(1)}$ sub-algebra generated by $J$. Note that, while
operators corresponding to $\widehat{B},\mathcal{D}$ and
$\bar{\mathcal{D}}$-type multiplets are guaranteed to be Virasoro
primaries \cite{Beem:2013sza}, those corresponding to
$\hat{\mathcal{C}}$-type multiplets are not. Therefore, this result is
already non-trivial. The fact that they are primary operators implies
that their non-vanishing OPEs with $J$ and $T$ are the following:
\begin{align}
 J(z)B^\pm(0) \sim \pm\frac{2B^\pm}{z}~&,\qquad T(z)B^\pm(0) \sim
 \frac{2B^\pm}{z^2} + \frac{\partial B^\pm}{z}~,
\\
J(z)D^{\pm I}(0) \sim \pm \frac{D^{\pm I}}{z}~&,\qquad T(z)D^{\pm
 I}(0)\sim \frac{2D^{\pm I}}{z^2} + \frac{\partial D^{\pm I}}{z}~, 
\\
J(z)\bar{D}^\pm_I(0) \sim \pm \frac{\bar{D}^\pm_I}{z}~&,\qquad
 T(z)\bar{D}^\pm_I(0) \sim \frac{2\bar{D}^\pm_I}{z^2} + \frac{\partial
 \bar{D}^{\pm}_I}{z}~,
\\
T(z)X(0) \sim \frac{3X}{z^2} + \frac{\partial X}{z}~,\qquad
 T(z)\bar{X}(0) &\sim \frac{3\bar{X}}{z^2} + \frac{\partial
 \bar{X}}{z}~,\qquad T(z)C_A(0) \sim \frac{3C_A}{z^2} + \frac{\partial C_A}{z}~&.
\end{align}

The OPEs of $B^\pm,D^{\pm I},\bar{D}^\pm_I,X,\bar{X}$ and $C_A$ are more
non-trivial. We evaluate them to find that the only non-vanishing OPEs of
these operators are the following. First, the OPEs of dimension-two
operators are
\begin{align}
	B^{+}(z)B^{-}(0)&\sim\frac{72}{z^4}-\frac{72 J}{z^3}
	+\frac{-16T-36\partial J+32J^2 }{z^2}\nonumber \\
	&\qquad +\frac{-20\partial^2 J-8{J^3}+16\partial
		J^2-8\partial T+16{TJ}}{z}\,, 
\end{align}
\begin{align}
	B^{\pm}(z) D^{\mp I }(0)
	\sim\frac{6D^{\pm I}}{z^2}
	\mp\frac{4{D^{\pm I}J}}{z}
	\,, &\quad 
	B^{\pm}(z) \bar{D}^{\mp}_{I}(0)
	\sim\frac{6\bar{D}^{\pm}_{I}}{z^2}
	\mp\frac{4{\bar{D}^{\pm}_{I}J}}{z}\,,\\
	D^{+ I}(z)D^{- J}(0)
	\sim \frac{- 2 \epsilon^{IJ} X}{z}\,, &\quad 	
	\bar{D}^{+}_{I}(z)\bar{D}^{-}_{J}(0)
	\sim \frac{- 2\epsilon_{IJ}\bar{X}}{z}\,, 
\end{align}
\begin{align}
	D^{\pm I}(z)\bar{D}^{\pm}_{J}(0)
	\sim \delta^{I}_{J}\left(
	\frac{2 B^{\pm}}{z^2}
	+\frac{\partial B^{\pm}}{z}				
	\right)\,, 
\end{align}
\begin{align}
			D^{\pm I}(z)\bar{D}^{\mp}_{J}(0)	
			&\sim
			\delta^{I}_{J}\left(
			\frac{24}{z^4}
			\mp\frac{12J }{z^3}+\frac{(2 J^2 -4T \mp 6\partial J)}{z^2}\right)\nonumber \\
			&\qquad +
			\delta^{I}_{J}\frac{
				\partial\left(J^2-2 T \right)\pm \left(2{TJ}-3\partial^{2}J\right) }{z}\pm\frac{ (\sigma^{A})^{I}{}_{J}C_{A}		}{z}~\,,
			\label{eq:dim2DD}
\end{align}
where $\epsilon_{IJ}$ and $\epsilon^{IJ}$ are anti-symmetric tensors such that
$\epsilon^{\uparrow\downarrow} =\epsilon_{\downarrow\uparrow} = 
1$, and $(\sigma^A)^I{}_J$ is the $(I,J)$-element of the Pauli matrix $\sigma^A$. 

The OPEs of dimension-two and three operators are
			\begin{align}
				B^{\pm}(z)X(0)
				&\sim \frac{\pm \epsilon_{IJ} D^{\pm I}D^{\pm J}}{z}\,,\quad 
				B^{\pm}(z)\bar{X}(0)\sim \frac{\pm \epsilon^{IJ}\bar{D}^{\pm}_{ I}\bar{D}^{\pm}_{J}}{z}
				\\
				B^{\pm}(z)C^{A}(0)
				&\sim \frac{\pm 2(\sigma^{A})^{I}{}_{J} D^{\pm J}\bar{D}^{\pm}_{I}}{z}\,, 
			\end{align}
			\begin{align}
				X(z)\bar{D}^{\pm }_{I}(0)
				&\sim \epsilon_{IK}\left(
				\frac{\pm 6 D^{\pm K}}{z^3}
				+\frac{ 2\left(JD^{\pm K} \pm 2 \partial D^{\pm K}\right)}{z^2}
			 +\frac{2\left(
				J\partial D^{\pm K}
				\pm \left(\partial^{2}D^{\pm K}-TD^{\pm K}\right)
				\right)}{z}\right)\\
					\bar{X}(z)D^{\pm I}(0)
					&\sim -\epsilon^{IK}\left(				
					\frac{\pm 6 \bar{D}^{\pm}_{K}}{z^3}
					+\frac{ 2\left( J\bar{D}^{\pm}_{K} \pm 2 \partial \bar{D}^{\pm}_{K}\right)}{z^2}+\frac{2\left(
						J\partial \bar{D}^{\pm}_{K}
						\pm \left(\partial^{2}\bar{D}^{\pm}_{K}-T\bar{D}^{\pm}_{K}\right)
						\right)}{z}\right)
\\
				C^{A}(z)D^{\pm I}(0)
				&\sim (\sigma^{A})^{I}{}_{K}
				\left(	\frac{\pm 6 D^{\pm K}}{z^3}
				+\frac{2(JD^{\pm K} \pm 2 \partial D^{\pm K})}{z^2}+\frac{2\left(
						J\partial D^{\pm K}
					\pm \left(\partial^{2}D^{\pm K}-TD^{\pm K}\right)
					\right)}{z}\right)\,,\\
				C^{A}(z)\bar{D}^{\pm }_{I}(0)
				&\sim (\sigma^{A})^{K}{}_{I}
				\left(	\frac{\pm 6 \bar{D}^{\pm}_{K}}{z^3}+\frac{2
					(J\bar{D}^{\pm}_{K} \pm 2\partial \bar{D}^{\pm}_{K})}{z^2}+\frac{2\left(
						J\partial \bar{D}^{\pm}_{K}
					\pm \left(\partial^{2}\bar{D}^{\pm}_{K}-T\bar{D}^{\pm}_{K}\right)
					\right)}{z}\right)\,.
			\end{align}

Finally, the OPEs of dimension-three operators are
			\begin{align}
				X(z) \bar{X}(0) &\sim -\frac{72}{z^6}+\frac{4\left( J^2+4T \right)}{z^4}
				+\frac{2 \partial \left( J^2+4T \right)}{z^3}\nonumber \\
				&\qquad -\frac{2}{z^2}
				\left(
				2T^2+2\partial (JT)-\partial^3 J-3\partial^2T+J\partial^2J 
				-D^{+ I}\bar{D}^{-}_{I}	 \right)\nonumber \\
				&\qquad -\frac{1}{z}\partial \left(
				2T^2+2\partial (JT)-\partial^3 J-\frac{7}{3}\partial^2 T+\frac{1}{3}\left(\partial J\right)^2
				+\frac{4}{3}J\partial^2 J
				\right)\nonumber \\
				&\qquad  +\frac{1}{z}\left(
				\partial D^{+ I}\bar{D}^{-}_{I}	 	-\bar{D}^{+}_{I}\partial D^{- I}	 \right)\,, 
			\end{align}
		\begin{align}
				C^{A}(z)X(0) \sim
				-\frac{2(\sigma^{A})_{IJ}D^{+ I} \partial D^{- J} }{z}
				\,, \quad 
				C^{A}(z)\bar{X}(0) \sim
				-\frac{2(\sigma^{A})^{IJ}\bar{D}^{+}_{I}
					\partial \bar{D}^{-}_{J} }{z}
		\end{align}
	\newpage
			\begin{align}
			C^{A}(z) C^{B}(0)
			 &\sim 
			\delta^{AB}\Big\{
			\frac{144}{z^6}-\frac{8\left( J^2+4T \right)}{z^4}
			-\frac{4 \partial \left( J^2+4T \right)}{z^3}
			 \nonumber \\
			&\left.\qquad +\frac{4}{z^2}\left(
			2T^2+2\partial (JT)-\partial^3 J-3\partial^2T+J\partial^2J 
			-D^{+ I}\bar{D}^{-}_{I}	 \right)\right. \nonumber\\
			&\qquad +\frac{2}{z}\partial \left(
			2T^2+2\partial (JT)-\partial^3 J-\frac{7}{3}\partial^2 T+\frac{1}{3}\left(\partial J\right)^2
			+\frac{4}{3}J\partial^2 J	-D^{+ I}\bar{D}^{-}_{I}	\right)\Big\} \nonumber\\
			&\qquad -\frac{2i f^{AB}{}_{C}(\sigma^{C})^{I}{}_{J}\left(
				D^{+ J}\partial \bar{D}^{-}_{I}+\bar{D}^{+}_{I}	\partial D^{-J}
				\right)}{z}\,.
			\label{eq:cc}
			\end{align}
Note that the OPEs of generators that are not listed above all vanish. In particular, we find
$B^\pm(z)D^{\pm I}(0)\sim 0$, $D^{\pm I}(z)D^{\pm J}(0)\sim 0$ and
$\bar{D}^{\pm}_I(z)\bar{D}^{\pm}_J(0) \sim 0$. The vanishing of these
OPEs are necessary for our conjecture that those in Table \ref{tb:generators} is the complete set of
generators, since no operator composed of those
in Table \ref{tb:generators} can appear in these OPEs for dimensional and symmetry
reasons.

			\subsection{Consistency check with Schur index}
\label{subsec:check}

In sub-section \ref{subsec:BRST2}, we conjecture that
$\mathtt{S}$ is the complete set of generators of the chiral algebra
$\chi[\mathcal{T}_{C_{2,0}}]$ associated with the genus two
theory $\mathcal{T}_{C_{2,0}}$. Operators in $\mathtt{S}$ are
listed in Table \ref{tb:generators}. As discussed in sub-section
\ref{subsec:BRST2}, we have checked that every
$Q_\text{BRST}$-cohomology class of dimension less than or equal to six
is generated by those in  $\mathtt{S}$. Here, we perform a further non-trivial check of this
conjecture by computing the character of
$\chi[\mathcal{T}_{C_{2,0}}]$ up to higher dimensions.

The (normalized) vacuum character of $\chi[\mathcal{T}_{C_{2,0}}]$ is
defined by 
\begin{align}
 \mathcal{I}(q,a) :=
 \text{Tr}_{\chi[\mathcal{T}_{C_{2,0}}]}(-1)^F
 q^{L_0}a^{J_0}~,
\label{eq:char}
\end{align}
where $L_0$ and $J_0$ are the zero modes of $T$ and $J$ respectively,
and $q, a\in \mathbb{C}$ such that $|q|<1$ and $|a|=1$. This character
can be
computed order by order in $q$ as follows. First, if $\chi[\mathcal{T}_{C_{2,0}}]$
has no null operator, $\mathcal{I}(q,a)$ is given by
				\begin{align}
				P.E.\left[
				\frac{1}{1-q}\left(
				q+(a^2+1+a^{-2})q^{2}-\left(4a+4a^{-1}\right)q^{2}+
				5q^{3}
				\right)
				\right]\,,
				\label{eq:generatorsindex}
				\end{align}
where	 $P.E.[g(q,a)]:= \exp\left[\sum_{n=1}^\infty	\frac{1}{n}g(q^n,a^n)\right]$ for any function $g$. 
				 Indeed, the above expression counts all composite operators built out of $\partial^k\mathcal{O}$ for 
				 $\mathcal{O}\in\mathtt{S}$ and $k\geq 0$. 
				 However, $\chi[\mathcal{T}_{C_{2,0}}]$ contains many null operators, corresponding to 4d operator 
				 relations in 
				 $\mathcal{T}_{C_{2,0}}$. 
				 These null operators must be removed from the spectrum.
				  Therefore, the vacuum
				 character $\mathcal{I}(q,a)$ of
				 $\chi[\mathcal{T}_{C_{2,0}}]$
				 is obtained by subtracting the
				 contributions of these null operators
				 from \eqref{eq:generatorsindex}. This
				 can be done by identifying the null
				 operators order by order in $q$. We
				 list all the null operators up to
				 dimension nine in Appendix \ref{app:null}.

On the 4d side, the vacuum character $\mathcal{I}(q,a)$ is identified with the Schur index
defined by 
\begin{align}
 \mathcal{I}_\text{Schur}(q,a) := \text{Tr}_{\mathcal{H}}(-1)^F
 q^{E-R}a^{f}~, 
\label{eq:Schur}
\end{align}
where $\mathcal{H}$ is the space of local operators in
$\mathcal{T}_{C_{2,0}}$, and $E, R$ and $f$ are the dimension,
$SU(2)_R$ charge and $U(1)_f$ charge of operators. Since
$\mathcal{T}_{C_{2,0}}$ has a Lagrangian description, one can
compute it via an integral formula \cite{Romelsberger:2005eg, Kinney:2005ej}. 

We have checked that \eqref{eq:Schur} agrees with \eqref{eq:char} up to
$\mathcal{O}(q^9)$. Combined with our check of the BRST cohomology classes
up to dimension six, this gives a strong evidence for our conjecture
that $\mathtt{S}$ is the
complete set of generators of $\chi[\mathcal{T}_{C_{2,0}}]$.

\section{Automorphisms and new $SU(2)$}
\label{sec:auto}

In this section, we discuss automorphisms of
$\chi[\mathcal{T}_{C_{2,0}}]$, based on its OPEs we identified
in the previous section. We will see that there exists
an unexpected $SU(2)$ automorphism sub-group in addition to
those associated with 4d flavor and $U(1)_r$ symmetries.

\subsection{Expected automorphisms}

Let us first recall that 4d flavor $U(1)_f$ symmetry gives rise to an $\widehat{\mathfrak{u}(1)}$  current, whose 
zero-mode $J_0$ gives a
$U(1)$ automorphism sub-group of
$\chi[\mathcal{T}_{C_{2,0}}]$. The $U(1)_f$ charges of the
generators are shown in Table \ref{tb:generators}. 

The charge conjugate
for this $U(1)_f$ leads to a $\mathbb{Z}_2$ automorphism under which
\begin{align}
J &\to -J\,,\qquad B^\pm \to -B^{\mp}\,,\qquad D^{\pm \uparrow} \to \pm i
 D^{\mp \downarrow}\,,\qquad D^{\pm \downarrow} \to \pm i
 D^{\mp \uparrow}\,,
\nonumber\\
\bar{D}^{\pm}_{\uparrow} &\to \pm i \bar{D}^{\mp}_{\downarrow}\,,\qquad
 \bar{D}^{\pm}_{\downarrow} \to \pm i \bar{D}^{\mp}_{\uparrow}~, \qquad
 C_1 \to -C_1~,
\end{align}
with the other generators
kept fixed. This $\mathbb{Z}_2$ corresponds to $\phi^{a \pm} \to \pm i
\phi^{a\mp}$, which preserves \eqref{eq:SB}  and $Q_\text{BRST}$, and
therefore is an automorphism of $\chi[\mathcal{T}_{C_{2,0}}]$.

There is also another $\mathbb{Z}_2$ automorphism corresponding to
exchanging two small $\mathcal{N}=4$ super Virasoro algebras. Under this
$\mathbb{Z}_2$, the generators of
$\chi[\mathcal{T}_{C_{2,0}}]$ transform
as 
\begin{align}
 B^\pm &\to -B^{\pm}\,,\qquad 
 D^{\pm\uparrow} \to \mp i D^{\pm\downarrow}~, \qquad 
D^{\pm \downarrow} \to \pm i D^{\pm \uparrow}\,,
\\
\bar{D}^{\pm}_{\uparrow} &\to \mp i\bar{D}^{\pm}_{\downarrow}\,\qquad 
 \bar{D}^{\pm}_{\downarrow} \to \pm i\bar{D}^{\pm}_{\uparrow}\,,\qquad
 C_1 \to -C_1\,,\qquad C_3 \to -C_3\,,
\end{align}
with the other generators kept fixed. The combined action of the above
two $\mathbb{Z}_2$ corresponds to exchanging the left and right sides of
the left quiver in Fig.~\ref{fig:quiver1}. Note that $D^{\pm I}$ and
$\bar{D}^\pm_I$ are eigenstates of this combined
$\mathbb{Z}_2$-action, where $D^{\pm \uparrow}$ and
$\bar{D}^{\pm}_{\uparrow}$ have eigenvalue $+1$ while $D^{\pm\downarrow}$
and $\bar{D}^\pm_{\downarrow}$ have eigenvalue $-1$.

As reviewed in sub-section
\ref{subsec:general}, the 4d $U(1)_r$
symmetry of $\mathcal{T}_{C_{2,0}}$ also gives rise to a $U(1)$
automorphism of $\chi[\mathcal{T}_{C_{2,0}}]$. The $U(1)_r$ charges of the chiral algebra generators are shown in
Table \ref{tb:generators}. It is straightforward to see that the OPEs
shown in sub-section \ref{subsec:OPEs} are all invariant under
$U(1)_r$.

\subsection{New $SU(2)$ symmetry}
\label{subsec:newsym}

In addition to the above $U(1)_f\times U(1)_r\times (\mathbb{Z}_2)^2$
automorphism sub-group expected from the construction, the chiral algebra $\chi[\mathcal{T}_{C_{2,0}}]$
turns out to have an unexpected action of $SU(2)$ under which 
			 \begin{enumerate}
			 	\item $D^{\pm I}$ and $\bar{D}_{\pm I}$
				      transform as doublets, with the
				      upper and lower indices $I$ being the
				      fundamental and anti-fundamental
				      indices, respectively,
			 	\item $C^{A}$ transform as a triplet with $A$ being the
				      adjoint index,
			 	\item and all the other generators transform
				      as singlets.
			 \end{enumerate}
Indeed, it is straightforward to see that the OPEs shown in sub-section \ref{subsec:OPEs} are all
covariant under this action of $SU(2)$.

In contrast to $U(1)_f$ and $U(1)_r$, the above $SU(2)$ symmetry
does not correspond to any
conserved one-form current in four dimensions. 
This follows from the fact that every conserved
one-form current in a 4d $\mathcal{N}=2$ SCFT is either a flavor current or an R-symmetry current.
\footnote{Here, 
the flavor
symmetry of a 4d $\mathcal{N}=2$ SCFT is defined as a global symmetry that commutes
with $\mathcal{N}=2$ superconformal symmetry.
} 
It is clear that our new
$SU(2)$ is not associated with any 4d R-current.\footnote{The
only $SU(2)$ sub-group of the 4d R-symmetry group is $SU(2)_R$, which is
broken in the 4d/2d correspondence.} It is also clear that this $SU(2)$
does not correspond to a 4d flavor current, since the genus two theory
$\mathcal{T}_{C_{2,0}}$ has only one flavor current multiplet,
$\hat{\mathcal{B}_1}$, corresponding to $U(1)_f$. Therefore, the
above $SU(2)$ symmetry either corresponds to a 4d
global symmetry without conserved currents, or is an accidental symmetry
in two dimensions.\footnote{In the latter case, it can be an accidental
enhancement of a smaller 4d global symmetry. It would particularly
interesting to see if this $SU(2)$ is an accidental enhancement of an
abelian higher-form symmetry \cite{Gaiotto:2014kfa} in four
dimensions. Note that the above $SU(2)$ symmetry cannot be interpreted
as coming from
an ``$SU(2)$ higher form symmetry'' in four dimensions since every
higher form symmetry is abelian.}
It would be interesting to
study each of these possibilities further.

While its four-dimensional origin is still to be understood, we see that
this $SU(2)$ symmetry strongly constrains the OPEs of $\chi[\mathcal{T}_{C_{2,0}}]$. For instance, the $U(1)_f\times
U(1)_r\times (\mathbb{Z}_2)^2$ symmetry does not forbid $C_2(0)/z^3$
to arise in
the OPE of $X(z)\bar{X}(0)$, which is however prohibited by the
$SU(2)$ symmetry; an $SU(2)$ triplet just cannot appear in the
tensor product of singlets. There are indeed various terms in the OPEs of generators that are forbidden by the $SU(2)$ symmetry.

Note that, since 2d OPEs are determined by 4d OPEs, this $SU(2)$
symmetry also constrains the OPEs of Schur operators in the genus two theory
$\mathcal{T}_{C_{2,0}}$. In particular, every 4d operator relation involving Schur operators gives rise to a null operator in
$\chi[\mathcal{T}_{C_{2,0}}]$, and therefore must be in some
representation of $SU(2)$. Since $J$ and $B^\pm$ are
neutral under it, this $SU(2)$ acts trivially on the Higgs
branch. To see the effects of this $SU(2)$, one needs to look at Schur
operators with non-vanishing spins.\footnote{Recall that Schur operators
are in $\hat{\mathcal{B}}_R, \, \mathcal{D}_{R(0,j_2)},\,
\bar{\mathcal{D}}_{R(j_1,0)}$ or $\hat{\mathcal{C}}_{R(j_1,j_2)}$.  As
seen from Table \ref{list:shcurops}, the only scalar Schur operators are
those in
$\hat{\mathcal{B}}_R$, i.e., Higgs branch operators.} This means that the
above unexpected $SU(2)$ symmetry of
$\chi[\mathcal{T}_{C_{2,0}}]$ constrains the OPEs of {\it non-scalar} Schur operators in
$\mathcal{T}_{C_{2,0}}$. We leave a detailed study of these non-trivial
constraints for future work.

			\section{ Conclusions and discussions
}\label{sec:conclusion}
			
			 In this paper, we studied the chiral algebra $\chi[\mathcal{T}_{C_{2,0}}]$ associated with the $A_1$-type 
			 genus two class $\mathcal{S}$ theory $\mathcal{T}_{C_{2,0}}$. 
			 We focus on the weak coupling description of the theory corresponding to the left quiver in
			 Fig.~\ref{fig:quiver1}, and apply the BRST reduction reviewed in
			 sub-section \ref{subsec:BRST}. 
			 We found that (1) BRST cohomology classes up
			 to dimension six are all built out of generators listed in sub-section
			 \ref{subsec:list}, and (2) these generators form closed OPEs. Given
			 these facts, we conjecture that these are the complete set of generators
			 of the chiral algebra $\chi[\mathcal{T}_{C_{2,0}}]$. 
			 As mentioned at the end of sub-section \ref{subsec:BRST2}, similar
			 conjectures were made for various other theories, leading to consistent
			 results. We also check that our conjecture is perfectly consistent with the Schur
			 index of $\mathcal{T}_{C_{2,0}}$ up to $\mathcal{O}(q^9)$.
			 The OPEs of these generators are shown in sub-section \ref{subsec:OPEs}.

One important and remarkable consequence of our OPEs is that there
exists an unexpected $SU(2)$ automorphism sub-group of
$\chi[\mathcal{T}_{C_{2,0}}]$. As discussed in section
\ref{sec:auto}, this $SU(2)$ symmetry is not related to any conserved
one-form current in four dimensions, and therefore either corresponds to
a 4d symmetry without conserved current or is an accidental symmetry in
two dimensions. We found that this $SU(2)$ acts trivially on 2d
operators corresponding to 4d Higgs branch operators. Therefore, this
$SU(2)$ is a symmetry characterizing the OPEs of non-scalar Schur operators.

While there has been various progress about the associated chiral algebra of class $\mathcal{S}$, our work is the first step to 
understand the chiral algebras of class $\mathcal{S}$ at higher genera.
			 There are indeed many open problems related to 
			 this work in this direction: 
			\begin{itemize}

		\item What is the four-dimensional origin of the new $SU(2)$
		      symmetry that we discussed in sub-section
		      \ref{subsec:newsym}? As mentioned already, there
		      is no 4d conserved one-form current corresponding
		      to it.
One possible way to understand it is to see how this $SU(2)$ symmetry emerges in the 
localization computation studied in \cite{Pan:2017zie,Pan:2019bor}.

				\item What is the chiral algebra of class $\mathcal{S}$ associated with a Riemann surface of higher genus? 
				It would be particularly interesting to see if there is a non-abelian   automorphism sub-group that does
				not  correspond to a 4d	conserved one-form current. 
				It would also be interesting to	consider a generalization to genus two theories arising from higher rank 6d 
				(2,0) theories.

				\item Is there any free field
				      realization of
				      $\chi[\mathcal{T}_{C_{2,0}}]$?
				      As shown in \cite{Bonetti:2018fqz, Beem:2019tfp, Beem:2019snk}, the
				      chiral algebras of a class of
				      4d $\mathcal{N}=2$ SCFTs have a
				      beautiful free field realization
				      that makes all the null operators
				      trivially vanishing. It would be
				      interesting to search for a
				      similar realization of
				      $\chi[\mathcal{T}_{C_{2,0}}]$,
				      which would also be useful for
				      solving the problem discussed in the
				      previous bullet.

				\item As shown in appendix D of
				      \cite{Beem:2017ooy}, the (normalized) Schur index of the genus two theory $\mathcal{T}_{C_{2,0}}$
				      satisfies a sixth-order modular linear
				      differential equation.  
				      This suggests that
				      $\chi[\mathcal{T}_{C_{2,0}}]$
				      has a null operator of dimension twelve, in addition to those listed in appendix \ref{app:null}. 
				      One can perform a further consistency check of our conjectured chiral algebra by identifying this null  
				      operator in it.

			\end{itemize}

		\section*{Acknowledgements}
		We thank Matthew Buican and Kazunobu Maruyoshi for illuminating discussions. Most of our computations are 
		done with the Mathematica package
		\verb|OPEdefs| provided by K.~Thielemans
		\cite{Thielemans:1991uw, Thielemans:1994er} to whom the authors are grateful.
		The work of K.~K is supported by JSPS KAKENHI Grant
		Number JP18J22009. T.~N.’s research is partially supported by
JSPS Grant-in-Aid for Early-Career Scientists 18K13547.

			\appendix
			\section{S-duality equivalence of Schur index}\label{app:index}

					As mentioned in section
				\ref{sec:gesnu2}, $\TT_{C_{2,0}}$ has
				two weak coupling Lagrangian
				descriptions corresponding to the quiver diagrams shown in Fig.~\ref{fig:quiver1}.
				These two quiver descriptions are expected to be related by S-duality, which implies the Schur
				indices computed via these quivers are equivalent.
				This equivalence has been checked in the case of vanishing flavor
				fugacity in \cite{Gadde:2011ik}.
				In this appendix, we will extend it to the case of non-vanishing flavor fugacity.
				Note that, since the flavor $U(1)_f$ symmetry of this theory is not
				visible in its class $\mathcal{S}$ construction, this extension
				does not immediately follow from the class $\mathcal{S}$ interpretation of S-duality.

			First, we focus on the left quiver diagram in Fig.~\ref{fig:quiver1}. Let $a$ be a $U(1)_{f}$ flavor fugacity. The 
			Schur index of $\TT_{C_{2,0}}$ is evaluated as 
			\begin{align}
			\mathcal{I}_1(q;a) &= \oint_{|x_i|=1}\left(\prod_{k=1}^3 \frac{dx_k}{2\pi ix_k}\Delta(x_k)\mathcal{I}_\text{vect}(q;x_k)\right)
			\mathcal{I}_\text{fund}(q;x_2,a)\mathcal{I}_\text{adj}(q;x_1,x_2)\mathcal{I}_\text{adj}(q;x_3,x_2)~,
			\label{eq:preI1}
			\end{align}
			where the contour integrations are taken over $|x_{i}|=1$, $\Delta(x_{k}):=\half (1-x_{k}^2)(1-x_{k}^{-2})$ is the factor arising from the Harr measure of $SU(2)_{k}$ gauge group, and 
			\begin{align}
				 \mathcal{I}_\text{vect}(q;x) &:= P.E.\left[\frac{-2q}{1-q}\left(x^2
				+1 + x^{-2}\right)\right]~,
				\\
				\mathcal{I}_\text{fund}(q;x,a) & :=
				P.E.\left[\frac{q^{ \half }}{1-q}(x+x^{-1})(a+a^{-1})\right]~,
				\label{eq:fund}
				\\
				\mathcal{I}_\text{adj}(q;x,a) &:= P.E.\left[\frac{q^{ \half }}{1-q}(x^2+1+x^{-2})(a+a^{-1})\right]~,
			\end{align}
			are respectively the index contributions from an $SU(2)$ vector multiplet, a
			fundamental hypermultiplet, and an adjoint hypermultiplet. Here, we
			used the plethystic exponential defined by
			\begin{align}
				P.E.[g(q;x_1,\cdots,x_k)]:= \exp\left[\sum_{n=1}^\infty
				\frac{1}{n}g(q^n;x_1^n,\cdots,x_k^n)\right]\,,
			\end{align}
			 for arbitrary function $g$ of fugacities.
			
			On the other hands, the right quiver diagram in Fig.~\ref{fig:quiver1} indicates the index of $\TT_{C_{2,0}}$ is 
			evaluated as
			\begin{align}
			\mathcal{I}_2(q;b) =
			\oint_{|x_i|=1}\left(\prod_{k=1}^3\frac{dx_k}{2\pi i
				x_k}\Delta(x_k)\mathcal{I}_\text{vect}(q;x_k) \right)\prod_{s=\pm1
			}\mathcal{I}_\text{tri-fund}^\text{half}(q;x_1,x_2,x_3,b^s)~,
			\label{eq:I2}
			\end{align}
			where $b$ is a fugacity for the $U(1)_{f}$ flavor symmetry, and
			\begin{align}
			\mathcal{I}_\text{tri-fund}^\text{half}(q;x_1,x_2,x_3,b):=
			P.E.\left[\frac{bq^{ \half }}{1-q}\prod_{k=1}^3(x_k+x_k^{-1})\right]\,,
			\label{eq:tri}
			\end{align}
			is the index contributions from trifundamental
			hypermultiplet. The relation between $a$ and $b$
			will be clear below.

When the flavor fugacity is turned off, the equivalence of
\eqref{eq:preI1} and \eqref{eq:I2} was shown in
\cite{Gadde:2011ik}. Indeed, it was shown in \cite{Gadde:2011ik} that
			\begin{align}
			\mathcal{I}_\text{tri-fund}^\text{half}(q;x_1,x_2,x_3,1) &=
			(q^2;q)\left[\prod_{k=1}^3\mathcal{I}_\text{vect}(q;x_k)\right]^{-\half}
			\!\!\sum_{R:\,\text{irreps of }				\mathfrak{su}(2)}
			\!\!\frac{\chi_R(x_1)\chi_R(x_2)\chi_R(x_3)}{[\text{dim}\,R]_q}~,
			\label{eq:tri1}
			\end{align}
			where $R$ runs over irreducible representations
 of $\mathfrak{su}(2)$, $(x;q):=\prod^{\infty}_{k=0}(1-x q^k)$,
 $[n]_q :=\frac{(q^{\frac{n}{2}}-q^{-\frac{n}{2}})}{(q^{\half}-q^{-\half})}$,
			and $\chi_R(x):= (x^{\text{dim}\,R} -
			x^{-\text{dim}\,R})/(x-x^{-1})$. Using this
			expression together with 
			\begin{align}
			\mathcal{I}_\text{fund}(q;x_2,1)\mathcal{I}_\text{adj}(q;x_1,x_2)
			\mathcal{I}_\text{adj}(q;x_3,x_2)
			= \mathcal{I}_\text{tri-fund}^\text{half}(q;x_1,x_1,x_2,1)\mathcal{I}_\text{tri-fund}^\text{half}(q;x_3,x_3,x_2,1)\,,
			\end{align}
and 
			\begin{align}
			\int_{|x|=1} \frac{dx}{2\pi i x}\Delta(x) \chi_{R_1}(x)\chi_{R_2}(x) =
			\delta_{R_1R_2}\,,
			\label{eq:identity1}
			\end{align}
one can show that the equivalence 
			\begin{align}
				\mathcal{I}_{1}(q;1) =\mathcal{I}_{2}(q;1) &= \sum_{R} \frac{(q^2;q)}{([\text{dim}\,R]_q)^2}~,
			\end{align}
in the case of $a=b=1$. 

Our aim in this appendix is to generalize the above proof to the case of
$a,b\neq 1$. First, by comparing the first few terms of $\mathcal{I}_1(q;a)$
and $\mathcal{I}_2(q;b)$, we see that the collect identification of the flavor fugacity is
			\begin{align}
				a = b ^ 2\,.
			\end{align}
Below, we rewrite $\mathcal{I}_1$ and $\mathcal{I}_2$ to show that
$\mathcal{I}_1(q;b^2) = \mathcal{I}_2(q;b)$.

			\subsection{Rewriting $\II_{1}$}
			
			From the identity
			\begin{align}
			\mathcal{I}_\text{adj}(q;x,a) &= P.E.\left[\frac{-q^{ \half }}{1-q}(a+a^{-1})\right]\times 
			\mathcal{I}_\text{tri-fund}^\text{half}(q;x,x,a,1)~,
			\end{align}
			and \eqref{eq:fund}, we see that 
			\begin{align}
			\mathcal{I}_1(q;b^2) 
			&= \oint_{|x_i|=1}\left(\prod_{k=1}^3 \frac{dx_k}{2\pi ix_k}\Delta(x_k)\mathcal{I}_\text{vect}(q;x_k)\right)
			P.E.\left[\frac{q^{ \half }}{1-q}(x_2+x_2^{-1})(b^2+b^{-2}-2)\right]
			\nonumber\\
			&\qquad\qquad\qquad \times
			\mathcal{I}^\text{half}_\text{tri-fund}(q;x_1,x_1,x_2)\mathcal{I}^\text{half}_\text{tri-fund}(q;x_3,x_3,x_2)~.
			\end{align}
			Using the identity \eqref{eq:identity1}, we can also rewrite it as
			\footnote{\begin{align*}
				\sum_{n=1}
				\frac{\chi_{n}(x)}{[\text{dim}\,n]_q}
				&=\frac{q^{-\half}-q^{\half}}{x-x^{-1}}\sum_{n=1}
				\left(
				\frac{q^{\frac{n}{2}}(x^n-x^{-n})}{1-q^n}
				\right)=\frac{1-q}{x-x^{-1}}\sum_{m=0}\left(
				\frac{q^{\half+m}x}{1-q^{\half+m}x}-\frac{q^{\half+m}x^{-1}}{1-q^{\half+m}x^{-1}}
				\right)\\
				&=\sum_{k=0}\frac{(1-q)q^k}{(1-q^{\half+k}x)(1-q^{\half+k}x^{-1})}
				\end{align*}}
			\begin{align*}
			\mathcal{I}_1(q;b^2) 
			&= (q^2;q)^2\oint_{|x|=1}\frac{dx}{2\pi i x}\Delta(x)
			P.E.\left[\frac{q^{ \half }}{1-q}(b^2+b^{-2}-2)(x+x^{-1})\right]
			\left(\sum_{n=1}
			\frac{\chi_{n}(x)}{[\text{dim}\,n]_q}\right)^2\\
			&=\oint_{|x|=1}\frac{dx}{2\pi i x}(1-x^2)f(x)\left(
			\sum_{m=0}\frac{q^m}{(1-q^{\half+m}x)(1-q^{\half+m}x^{-1})}\right)^2\,,
			\end{align*}
			where 
			\begin{align*}
			f(x)=(q;q)^2
			P.E.\left[\frac{q^{ \half }}{1-q}(b^2+b^{-2}-2)(x+x^{-1})\right]
			=\frac{(q;q)^2(q^{\half}x;q)^2(q^{\half}x^{-1};q)^2}{\prod_{s_{1},s_{2}=\pm}(q^{\half}b^{2s_{1}} 
			x^{s_{2}};q)}\,.
			\end{align*}
			The last line are rewritten for symmetry $f(x^{-1})=f(x)$.

			We now evaluate the residues of the contour integral. 
			The poles of the integrand are at $x=q^{\half+k}b^{\pm 2}$ for $k\geq 0$.\footnote{
				There is no pole at $x=0$, 
			since $f(0)$ takes finite value and the summation part in the integrand is zero in the limit $x \to 0\,.$}
			First, we evaluate the residue at the pole at $
			x=q^{\half+k}b^{2}$. The residue involves the
			following factor from the Pochhammer symbol
			\begin{align*}
				\frac{(x-q^{\half+k}b^{2})}{x(q^{\half+k}b^{2}x^{-1};q)}\eval{}_{x=q^{\half+k}b^{2}}
				=	\prod_{t=0}^{k-1}\frac{1}{(1-q^{t-k})}	
				\prod_{t'=k+1}^{\infty}\frac{1}{(1-q^{t'-m})}=\frac{(-1)^{k}q^{\half k(k+1)}(q^{k+1};q)}{(q;q)^2}\,,
			\end{align*}
and therefore we find
			\begin{align}
				\frac{(x-q^{\half+k}b^{2})}{x}f(x)\eval{}_{x=q^{\half+k}b^{2}}
				&=\frac{(-1)^{k}q^{\half k(k+1)}(q^{k+1}b^2;q)^2(q^{-k}b^{-2};q)^2}{(q^{k+1}b^4;q)(q^{-k}b^{-4};q)}\nonumber \\
				&~~=\frac{1-b^{-2}}{1+b^{-2}}\frac{(b^{2}q;q)^2(b^{-2}q;q)^2}{(b^{4}q;q)(b^{-4}q;q)}\,.
			\end{align}
			 In the last equality, we used the formula 
			\begin{align}
			(aq^{k+1};q)(a^{-1}q^{-k};q)=(-a^{-1})^kq^{-\frac{k(k+1)}{2}} (aq;q)(a^{-1};q)\,.
			\end{align} 
			From the above calculations, we see that the residue at $ x=q^{\half+k}b^{2}$ is
			\begin{align}
				\frac{1-b^{-2}}{(1-b^4 q^{2k+1})(1+b^{-2})}\frac{(b^{2}q;q)^2(b^{-2}q;q)^2}{(b^{4}q;q)(b^{-4}q;q)}
				\left(\sum_{m=0}\frac{(1-b^4q^{2m+1})q^k}{(1-q^{k+m+1}b^2)(1-q^{m-k}b^{-2})}\right)^2\,.
			\end{align}
			Note here that one can further rewrite the sum over $m$ as
			\begin{align}		 	
			\sum_{m=0}\frac{(1-b^4 q^{2k+1})q^m}{(1-b^2q^{k +m+1})(1-b^{-2}q^{m-k})}
			&=\sum_{m=0}\left(
			\frac{-b^4q^{2k+m+1}}{1-b^2q^{k +m+1}}-\frac{b^2q^k}{1-b^2q^{k-m}}
			\right)\nonumber \\
			&=-b^2q^{k}\left(k+\frac{1}{1-b^2}+
			\sum_{\ell=1}\frac{(b^{2\ell}-b^{-2\ell})q{^\ell}}{1-q{^\ell}}
			\right)\,.\label{eq:I1formula}
			\end{align}
Combining the above result with the residues at 
 			$ x=q^{\half+k}b^{-2}$, we finally get the
 			following expression for $\mathcal{I}_1(q;b^2)$:
				\begin{align}
			\mathcal{I}_1(q;b^2) 
			&=\frac{1-b^{2}}{1+b^{2}}\frac{(qb^2;q)^2(qb^{-2};q)^2}{(qb^4 ;q)(qb^{-4};q)}\nonumber \\ 
			& \times \left(\sum_{m=0}\frac{-b^4q^{2m}}{(1-b^4 q^{2m+1})}
			\left(m+\frac{1}{1-b^2}+\sum_{\ell=1}\frac{(b^{2\ell}-b^{-2\ell})q{^\ell}}{1-q{^\ell}}\right)^2
			-(b\rightarrow b^{-1})\right)\,.\label{eq:I1last}
			\end{align}

			\subsection{Rewriting $\II_{2}$} 
			Let us now turn to $ \II_{2}(q;b)$.
			Using the identity
			\begin{align}
				\begin{aligned}
					&\mathcal{I}_\text{tri-fund}^\text{half}(q;x_1,x_2,x_3,b)\mathcal{I}_\text{tri-fund}^\text{half}(q;x_1,x_2,x_3,b)\\
					&\qquad=\mathcal{I}_\text{tri-fund}^\text{half}(q;x_1,bx_2,x_3,1)
					\mathcal{I}_\text{tri-fund}^\text{half}(q;x_1,b^{-1}x_2,x_3,1)
				\end{aligned}
			\end{align}
			and \eqref{eq:tri1}, we can more simplify $\II_{2}(q;b)$ as 
			\begin{align}
				&\II_{2}(q;b)\nonumber\\
				&\qquad=(q^2;q)^2\oint_{|x|=1}
					\frac{dx}{2\pi i x}(1-x^2)
					P.E.\left[\frac{q}{1-q}(b^2+b^{-2}-2)(x^{2}+x^{-2})\right]
				\sum_{R}\frac{\chi_{R}(bx
			 )\chi_{R}(b^{-1}x )}{[\text{dim} R]^2}\,.
\label{eq:I2-again}
			\end{align}
Note that the sum over $R$ can be rewritten as
			\begin{align}
			\sum_{R} \frac{\chi_{R}\left(b x\right) \chi_{R}\left(b^{-1} x\right)}{\left([\operatorname{dim} R]_{q}\right)^{2}}=\frac{x^2(1-q)^2}{(1-b^2x^2)(1-b^{-2}x^{2})}g(x)\,,
			\end{align}
where
			\begin{align}
				g(x):=\sum_{k=0}(k+1)q^{k}\left(
				\frac{x^2+x^{-2}-2q^{k+1}}{(1-x^2q^{k+1})(1-x^{-2}q^{k+1})}
				-\frac{b^2+b^{-2}-2q^{k+1}}{(1-b^2q^{k+1})(1-b^{-2}q^{k+1})}\right)\,.
			\end{align}
From this and 
			\begin{align}
			 P.E.\left[\frac{q}{1-q}(b^2+b^{-2}-2)(x^{2}+x^{-2})\right]
			 =\frac{(q;q)^2(qx^2;q)^2(qx^{-2};q)^2}{\prod_{s_1,s_2=\pm 1}(qb^{2s_{1}}x^{2s_2})}\,,
			\end{align}
we see that the integrand of \eqref{eq:I2-again} has a pole at $x= \pm b^{\pm
1}q^{\frac{k+1}{2}}$ for all $k \in \mathbb{Z}_{\geq 0 }$.
\footnote{For careful evaluation, we can see that $x=0,\pm b^{\pm}$ are also not poles. }

Note that the residue at $x= bq^{\frac{k+1}{2}}$ is evaluate as
			\begin{align}
				\frac{-(1-b^2)}{2(1+b^2)}\frac{(b^2;q)^2(b^{-2}q;q)^2}{(b^4q;q)(b^{-4}q;q)}
				\frac{b^2 q^{k+1}}{1-b^2 q^{k+1}}g( bq^{\frac{k+1}{2}})\,.
			\end{align}
Combining this and its cousin obtained by $b\to -b^{-1}$, we find
			 \begin{align}
			 \II_{2}(q;b)&=\frac{-(1-b^2)(b^2;q)^2(b^{-2}q;q)^2}{(1+b^2)(b^4q;q)(b^{-4}q;q)}\!\!
			 \sum_{k=0,m=0}\!\!\frac{(m+1)b^2 q^{k+m+1}}{1-b^2 q^{k+1}}
			 \frac{b^2q^{k+1}+b^{-2}q^{-(k+1)}-2q^{m+1}}{(1-b^2q^{k+m+2})(1-b^{-2}q^{k-m})}\nonumber\\
			 &\qquad +(b \to b^{-1})\,.
\label{eq:I2-further-again}
			 \end{align}
Since the sum over $k$ and $m$ is rewritten as\footnote{
Here we use 
			 	\begin{align}
			 	&\sum_{k=0}\frac{-(k+1)q^{k+1}}{q^{1+k}-b^2q^{m+1}}
			 	=\sum_{k=0}^{m}\frac{-(k+1)}{1-b^2q^{m-k}}+\sum_{k=0}^{\infty }
			 	 \frac{(k+1+m+1)b^{-2}q^{k+1}}{1-b^{-2}q^{k+1}}\\
			 	&\qquad =
			 	\sum_{k=0}^{m}\frac{k-(m+1)}{1-b^2q^k}+(m+1)\sum_{\ell =1}^{\infty } 
			 	\frac{b^{-2\ell}q^{\ell}}{1-q^{\ell}}+\sum_{\ell=1}\frac{b^{-2\ell}q^{\ell}}{(1-q^{\ell})^2}
			 	\end{align}
			 	and 
			 	\begin{align}
			 	&\sum_{k=0}^{m}\frac{1}{1-b^2q^k}
			 	=\frac{1}{1-b^2}+m+\sum_{\ell=1}\frac{b^{2\ell}q{^\ell}(1-q^{ml})}{1-q^{l}}\\
			 	&\sum_{k=1}^{m}\frac{k}{1-b^2q^k}
			 	=\half m(m+1)+\sum_{\ell=1}\frac{	b^{2\ell}(q^{l}-(m+1)q^{l(m+1)}+mq^{l(m+2)})}{(1-q{^\ell})^2}\,.
			 	\end{align}
		 		}
			 \begin{align}
				&\sum_{m,k=0}\frac{(m+1)b^2 q^{k+m+1}}{1-b^2 q^{k+1}}\left(
			 	\frac{-1}{q^{k+1}-b^2q^{m+1} }
			 			\right)
			 		+\sum_{\ell=1}\frac{b^{2\ell}q^{(m+2)\ell}-(b^{2\ell }+b^{-2\ell })q^{\ell}}{(1-q^\ell)^2}\nonumber\\
			 	&\qquad =\sum_{m=0}\frac{(m+1)b^2q^{m}}{1-b^2 q^{m+1}}\left(
			 	\half m +\frac{1}{1-b^2}+\sum_{\ell=1}\frac{(b^{2\ell} -b^{-2\ell})q{^\ell}}{1-q{^\ell}}
			 	\right)\,,
			 \end{align}
the expression \eqref{eq:I2-further-again} is further rewritten as	 
			 \begin{align}
			 \mathcal{I}_{2}(q ; b)&= \frac{1-b^2}{1+b^2}
			 \frac{(qb^2;q)^2(qb^{-2};q)^2}{(qb^4;q)(qb^{-4};q)}\nonumber\\
			 &\sum_{m=0}\left(
			 \frac{(m+1)b^2q^{m}}{1-b^2 q^{m+1}}\left(
			 \half m +\frac{1}{1-b^2}+\sum_{\ell=1}\frac{(b^{2\ell} -b^{-2\ell})q{^\ell}}{1-q{^\ell}}\right)-(b\rightarrow b^{-1})
			 \right)\,.
			 \label{eq:I2fin}
			 \end{align}
			 
			 \subsection{Proof of $\II_{1}=\II_{2}$ }

From \eqref{eq:I1last} and \eqref{eq:I2fin}, we see that proving
$\II_{1}=\II_{2}$ is equivalent to proving
			\begin{align}
				h_{1}(b)+h_{2}(b)=h_{1}(b^{-1})+h_{2}(b^{-1})
				\label{eq:toshow}\,,
			\end{align}
where
			\begin{align}
			h_{1}(b)&:=\sum_{m=0}^{\infty} 
			\left(\frac{b^{4} q^{2 m}}{1-b^{4} q^{2 
			m+1}}\left(m+\sum_{k=1}^{\infty}{\frac{b^2q^{k}}{1-b^2q^{k}}}
				-\sum_{k=0}^{\infty}\frac{b^{-2}q^{k}}{1-b^{-2}q^{k}}\right)^2\right)	\,,
\label{eq:foo}
 	\\		
			h_{2}(b)&:=\sum_{m=0 }^{\infty}\frac{(m+1)b^2q^{m}}{1-b^2q^{m+1}}	
			\left(\frac{m}{2}+\sum_{k=1}^{\infty}{\frac{b^2q^{k}}{1-b^2q^{k}}}
			-\sum_{k=0}^{\infty}\frac{b^{-2}q^{k}}{1-b^{-2}q^{k}}\right)\,.
			\end{align}

In the rest of this appendix, we prove \eqref{eq:toshow} by showing that both sides of
\eqref{eq:toshow} have the same poles and residues.
			Both sides of \eqref{eq:toshow} have poles only	at $b=\pm q^{\frac{\ell}{2}}, \ell \in \mathbb{Z} $.
			 Note that potential poles of the LHS at $b=\pm s
			q^{-\frac{\ell}{2}-\frac{1}{4}}$ for $s=1,i$,
			and those of the RHS at $b=\pm s
			q^{\frac{\ell}{2}+\frac{1}{4}}$ for 
 $\ell \in \mathbb{Z}_{\geq 0}$ have vanishing residues. 
 			For instance, the residue of the expression in the most inner bracket in
			\eqref{eq:foo}, at $b= s q^{-\frac{\ell}{2}-\frac{1}{4}}$, is
			\begin{align*}
			&\ell+\frac{1}{1-s^2q^{-\ell-\half}}+\sum_{k=1}
			\left(
			\frac{s^2q^{k-\ell-\half}}{1-s^2 q^{k-\ell-\half}}
			-\frac{s^2 q^{k+\ell+ \half }}{1-s^2q^{k+\ell+ \half }}
			\right)\\
			&~=
			\ell-\frac{s^2q^{\ell+\half}}{1-s^2q^{\ell+\half}}
			+\sum_{k=1}^{2\ell}
			\frac{s^2q^{k-\ell-\half}}{1-s^2q^{k-\ell-\half}}+
			\frac{s^2q^{\ell+\half}}{1-s^2 q^{\ell+\half}}	+
			\sum_{k=1}\left(
			\frac{s^2q^{k+\ell+\half}}{1-s^2 q^{k+\ell+\half} }
			-\frac{s^2 q^{k+\ell+ \half }}{1-s^2q^{k+\ell+ \half }}	
			\right)\\
			&~=0\,.
			\end{align*}
Therefore the expression in the most inner bracket is regular at
			$b=sq^{-\frac{\ell}{2}-\frac{1}{4}}$. This means
			that $h_1(b)$ has no pole at this point.
			Similarly, one can show that both sides of \eqref{eq:toshow} have no
			poles at $b=\pm s q^{-\frac{\ell}{2}- \frac{1}{4}}$ or $b= \pm s q^{\frac{\ell}{2}+\frac{1}{4}}$.
			Below, we show that the residues of both sides of \eqref{eq:toshow} at
			$b=\pm q^{\frac{\ell}{2}}$ are identical, for $\ell\in \mathbb{Z}$.

			\subsubsection{Poles and residues on both sides}

			Let us focus on the poles at $b= q^{\frac{\ell}{2}}$ for $\ell>0$. Its
			generalization to the other poles is straightforward.
			Note that these poles are second order poles. One can see that the coefficients of the most
			singular terms on both sides of
			\eqref{eq:toshow} are
			identical. Indeed, when $b\sim
			q^{\frac{\ell}{2}}$, 
			the LHS behaves as  
			\begin{align}
			h_1(b) + h_2(b) \sim \frac{1}{1-b^{-2}q^{\ell}}\sum_{m=0}^{\infty}\frac{q^{3\ell+2m}}{4(1-q^{2m+2\ell+1})}\,.
			\end{align}
			On the other hand, the RHS behaves as
			\begin{align}
			h_1(b^{-1}) + h_2(b^{-1}) 
			& \sim
			 \frac{1}{1-b^{-2}q^{\ell}}
			 \left(\sum_{m=0}^\infty \frac{q^{2m-\ell}}{4(1-q^{2(m-\ell)+1})}+\frac{\ell}{4}q^{\ell-1}\right)\\
			&=\frac{1}{1-b^{-2}q^{\ell}}\sum_{m=0}^\infty \frac{q^{3\ell+2m}}{4(1-q^{2(m+\ell)+1})}\,.
			\end{align}
Thus, the most singular terms on both sides agree.
		
			Let us next evaluate the residues on both sides. 
			First, we compute the residue on the LHS. The residue coming from $h_{1}(b)$ is evaluated as
			\begin{align*}
			&\sum_{m=0}^{\infty}
			\Bigg[\frac{-q^{2(m+\ell)}q^{\frac{\ell}{2}}}{1-q^{2 
			(m+\ell)+1}}\left(m+\sum_{k=1}^{\infty}{\frac{q^{k+\ell}}{1-q^{k+\ell}}}
			-\sum_{\substack{k=0\\[.5mm](k\neq \ell)}}^{\infty}\frac{q^{k-\ell}}{1-q^{k-\ell}}\right)
			+ \dv{b}\eval{\left(\frac{b^{4} q^{2 m}}{1-b^{4} q^{2 m+1}}
				\frac{q^{2\ell}}{(b+q^{\frac{\ell}{2}})^2}\right)}_{b=q^{\frac{\ell}{2}}}\Bigg]\\
			&~=q^{\frac{\ell}{2}}\sum_{m=0}^{\infty}
			\left(\frac{-mq^{2m}}{1-q^{2m+1}}
			+\frac{q^{2m}(3+q^{2m+1})}{4(1-q^{2m+1})^2}\right)
			+q^{\frac{\ell}{2}}\sum_{m=0}^{\ell-1}\left(
			\frac{mq^{2m}}{1-q^{2m+1}}-\frac{q^{2m}(3+q^{2m+1})}{4(1-q^{2m+1})^2}
			\right)\,,
			\end{align*}
			while, the residue coming from $h_{2}(b)$ is evaluated as 
			\begin{align}
			q^{\frac{\ell}{2}} \sum_{m=0}\frac{-(m+1)q^{m+\ell}}{2(1-q^{m+\ell+1})}
			=q^{\frac{\ell}{2}} \sum_{m=0}\frac{-(m-\ell +1)q^{m}}{2(1-q^{m+1})}+q^{\frac{\ell}{2}} \sum_{m=0}^{\ell-1}\frac{(m-\ell +1)q^{m}}{2(1-q^{m+1})}\,.
			\end{align}
			Combining the above two, we see that the residue of the LHS of
			\eqref{eq:toshow} is
			\begin{align}
			&q^{\frac{\ell}{2}}\sum_{m=0}^{\infty}
			\left(
			\frac{-mq^{2m}}{1-q^{2m+1}}+\frac{q^{2m}(3+q^{2m+1})}{4(1-q^{2m+1})^2}
			-\frac{(m+1-\ell )q^{m}}{2(1-q^{m+1})}\right)\nonumber \\
			&\qquad +q^{\frac{\ell}{2}}\sum_{m=0}^{\ell-1}\left(
			\frac{mq^{2m}}{1-q^{2m+1}}-\frac{q^{2m}(3+q^{2m+1})}{4(1-q^{2m+1})^2}
			+\frac{(m+1-\ell )q^{m}}{2(1-q^{m+1})}
			\right)\,.
			\label{eq:resleft}
			\end{align}

We next turn to the RHS of \eqref{eq:toshow}. 
The residue coming from $h_{1}(b^{-1})$	is 
				\begin{align}
			&\sum_{m=0}\frac{q^{2(m-\ell)}q^{\frac{\ell}{2}}}{1-q^{2(m-\ell)+1}}
			\left(m +\sum_{k=1,k\neq \ell}^{2\ell-1}{\frac{q^{k-\ell}}{1-q^{k-\ell}}}\right)
			+\dv{b}\frac{q^{2(m+\ell)}}{(b^4-q^{2m+1})(b+q^{\frac{\ell}{2}})^2} \!\!\eval{}_{b=q^{\frac{\ell}{2}}}\\
			&~=q^{\frac{\ell}{2}}\sum_{m=0}\left(\frac{mq^{2m}}{1-q^{2m+1}}-\frac{q^{2m}(1+3q^{2m+1})}{4(1-q^{2m+1})^2}\right)
			+\sum_{m=0}^{\ell-1}\left(\frac{m}{q(1-q^{2m+1})}+\frac{(1-5q^{2m+1})}{4q(1-q^{2m+1})^2}\right)\,.
			\end{align}
			and that arising from $h_{2}(b^{-1})$ is
			\begin{align*}
			&\sum_{m+1\neq\ell}\frac{(m+1)q^{m-\frac{\ell}{2}}}{2(1-q^{m-\ell+1})}
			+\frac{l}{2}q^{\frac{\ell}{2}-1}\left( \frac{l-1}{2}
			+\sum_{k=1,k \neq \ell}^{2\ell-1}\frac{q^{k-\ell}}{1-q^{k-\ell}}\right)
			+\dv{b}\left(	\frac{lq^{2\ell -1}}{(b+q^{\frac{\ell}{2}})^2}\right)\eval{}_{b=q^{\frac{\ell}{2}}}\\
			&=q^{\frac{\ell}{2}}\sum_{m=0}^{\ell-1}\frac{(m+1-\ell )}{2q(1-q^{m+1})}
			+q^{\frac{\ell}{2}}\sum_{m=0}\frac{(m+\ell+1)q^{m}}{2(1-q^{m+1})}
			-\frac{\ell^2}{4}q^{\frac{\ell}{2}-1}\,.
			\end{align*}
			Therefore, the residue of the RHS of \eqref{eq:toshow} at $b=q^{\frac{\ell}{2}}$ is evaluated as  
			\begin{align}
			&q^{\frac{\ell}{2}}\sum_{m=0}\left(
			\frac{mq^{2m}}{1-q^{2m+1}}
			-\frac{q^{2m}(1+3q^{2m+1})}{4(1-q^{2m+1})^2}
			+\frac{(m+1+\ell)q^{m}}{2(1-q^{m+1})}\right)\nonumber \\
			&\qquad ~+q^{\frac{\ell}{2}}\sum_{m=0}^{\ell-1}
			\left(\frac{m}{q(1-q^{2m+1})}+\frac{(1-5q^{2m+1})}{4q(1-q^{2m+1})^2}+\frac{(m+1-\ell 
			)}{2q(1-q^{m+1})}\right)
			-\frac{\ell^2}{4}q^{\frac{\ell}{2}-1}\,.
			 \label{eq:resright}
			\end{align} 
					
			\subsubsection{Coincidence of the residues}

			We here show that the residue \eqref{eq:resleft} of the LHS of \eqref{eq:toshow} agrees with the residue 
			\eqref{eq:resright} of the RHS.

			Note first that the second line of \eqref{eq:resleft} minus that of
			\eqref{eq:resright} is simplified as
			\begin{align}
				-q^{\frac{\ell}{2}}\sum_{m=0}^{\ell-1}\frac{(6m-2\ell+3)}{4q} +
				\frac{\ell^2}{4}q^{\frac{\ell}{2}-1} = 0~.
			\end{align}
			Therefore, all we need to show is the
			 equivalence of the first lines of
			 \eqref{eq:resleft} and
			 \eqref{eq:resright}. Note that this is
			 equivalent to proving the identity
		\begin{align}
		\sum_{m=0}^\infty\left(\frac{2m q^{2m}}{1-q^{2m+1}}+\frac{(m+1)q^m}{1-q^{m+1}}\right)
		=\sum_{m=0}^\infty \frac{q^{2m}(1+q^{2m+1})}{(1-q^{2m+1})^2}\,.	
		\label{eq:indexfin1}
		\end{align}
Using the identities
		\begin{align}
		&\sum_{m=0}^\infty\frac{2m
		 q^{2m}}{1-q^{2m+1}}=\sum_{m=0}^\infty 
	 \left(\frac{2q^{m}}{(1-q^{2(m+1)})^2}+\frac{-2q^{2m}}{1-q^{2m+1}}\right)~,\\
		&		 \sum_{m=0}^\infty\frac{(m+1)q^m}{1-q^{m+1}}
		=\sum_{m=0}^\infty \left(\frac{q^{2m}}{(1-q^{2m+1})^2}+\frac{q^{2m+1}}{(1-q^{2(m+1)})^2}\right)~,
		\end{align}
we see that proving \eqref{eq:indexfin1} is equivalent to proving 
		\begin{align}
		\sum_{m=0}^\infty\left(\frac{q^{m}(2+q^{m+1})}{(1-q^{2(m+1)})^2}-\frac{2q^{2m}}{(1-q^{2m+1})^2}+\frac{q^{4m+1}}{(1-q^{2m+1})^2}\right)=0\,.
		\label{eq:indexfin2}
		\end{align}
Below, we show that \eqref{eq:indexfin2} indeed holds. To that end,
first note that 
		\begin{align}
		\sum_{m=0}^\infty\frac{2q^{2m}}{(1-q^{2m+1})^2} 
		&=\sum_{m=0}^\infty\left(\frac{2q^m}{(1-q^{m+1})^2}-\frac{2q^{2m+1}}{(1-q^{2(m+1)})^2}\right)\,,\\
		\sum_{m=0}^\infty \frac{q^{4m+1}}{(1-q^{2m+1})^2} 
		&=\sum_{m=0}^\infty\left(\frac{q^{2m+1}}{(1-q^{m+1})^2}-\frac{q^{4m+3}}{(1-q^{2(m+1)})^2}\right)\,.
		\end{align}
Using these identities, the LHS of \eqref{eq:indexfin2} is rewritten as
		\begin{align}
		&\sum_{m=0}^\infty\left(\frac{-q^{2m+1}(1+2q^{m+1})}{(1-q^{2(m+1)})^2}+\frac{q^{2m+1}}{(1-q^{m+1})^2}-
		\frac{q^{4m+3}}{(1-q^{2(m+1)})^2}\right)\,,
		\end{align}
which can be shown to vanish by a straightforward
calculation. Therefore, \eqref{eq:indexfin2} is an identity, which
completes our proof of the equivalence between \eqref{eq:resleft} and \eqref{eq:resright}.

			\section{Null operators of genus two chiral algebra}\label{app:null}

In this appendix, we list the null operators in the chiral algebra
			$\chi[\mathcal{T}_{C_{2,0}}]$ whose holomorphic
			dimension (which we denote by $h$) is less than or equal to nine. Interestingly, such null operators
			only exist at $h=4,5$ and $6$, up to
			composite operators involving them or their derivatives. Note that the
			absence of independent null operators at $h=7,8$ and $9$
			does not mean the absence of such operators at
			$h\geq 10$. Indeed, the modular linear
			differential equation studied in \cite{Beem:2017ooy} suggests an independent null operator
			involving $T^6$ at $h=12$. 
			It would be interesting to extend our results here to higher dimensions.

			 Below, we list these null operators as operator relations. 
			 We also classify these operator relations in
			 terms of the $U(1)_f$ charge, $f$, and the
			 $U(1)_r$ charge $r$. Note that every pair of
			 generators, $X$ and $Y$, satisfies the
			 following trivial operator relation
\begin{align}
 YX -(-1)^{|X||Y|} XY = \sum_{n\geq 1} \frac{(-1)^n}{n!} \partial^{n} [XY]_n~,
\end{align}
where $|\mathcal{O}|=0$ or $|\mathcal{O}|=1$ when $\mathcal{O}$ is
bosonic or fermionic respectively, and $[XY]_n$ are operators such that
$X(z)Y(0) \sim
\sum_{n\geq 1}[XY]_{n}(0)/z^n$ \cite{Thielemans:1994er}. Since these
relations just follow from the definition of the normal ordered product,
we do not list them below.\footnote{In our computation of the character in
section \ref{subsec:check}, these operator relations are taken into
account by counting only one of $XY$ and $YX$. Indeed,
the expression \eqref{eq:generatorsindex} assumes that $XY$ and $YX$ are
linearly dependent and therefore counts only one of them as an
independent operator.}
			 
			 \subsection{Dim 4}
			 The null operators at dimension four are as follows. 
			 
\subsubsection*{Nulls with $f=r=0$}
First, null operator relations  with $f=r=0$ are
\begin{align}
		{D^{+I}\bar{D}^{-}_{J}}+{\bar{D}^{+}_{J}D^{- I}}= 0\,,
	\\
	\partial C^{A}={JC^{A}}+(\sigma^{A})^{I}{}_{J}{D^{+ J}\bar{D}^{-}_{I}}\,,
\end{align}
			\begin{align}
				\left({B^{+}B^{-}}-J^4\right) 	+2\left(D^{+ I}\bar{D}^{-}_{I}\right)
				 &=4T^2 -6 \partial^2 T -8{J^2 T}+12\partial {(J T)} -4 \partial J^3
				 \nonumber\\
				 &\qquad +9(\partial J)^2 +14{ J\partial^2J }-5\partial^3 J \,.
				 \label{eq:higgsnull}
			\end{align}
			We see that, when we omit derivatives and generators
			corresponding to
			$\mathcal{D}_{R(0,\bj)},\,\bar{\mathcal{D}}_{R(j,0)}$
			and $\hat{\mathcal{C}}_{R(j,\bj)}$, the last
		 operator relation \eqref{eq:higgsnull} is identical
			to the Higgs branch chiral ring relation
			\eqref{eq:higgrel} in four dimensions. The
			mixing with derivative operators and those
			arising from $\mathcal{D}_{R(0,\bj)},\,\bar{\mathcal{D}}_{R(j,0)}$
			or $\hat{\mathcal{C}}_{R(j,\bj)}$ is a common
			feature of associated chiral algebras.

\subsubsection*{Nulls with $f=\pm 1$ and $r=\pm \half $}
Second, the null operator relations for $f=\pm 1$ and $r=\pm  \half $ are 
			\begin{align}
			\partial^2 D^{\pm I}&=
			T D^{\pm I}\pm \partial (J D^{\pm I})
			+\half \left(B^{\pm}D^{\mp I }- J^2D^{\pm I} \pm\partial JD^{\pm I}\right)\,,
			 \nonumber \\
			\partial^2 \bar{D}^{\pm}_{I}&=
			T \bar{D}^{\pm}_{I} \pm \partial (J \bar{D}^{\pm}_{I})
			+\half \left(B^{\pm}\bar{D}^{\mp}_{I}- J^2\bar{D}^{\pm}_{I} \pm\partial J\bar{D}^{\pm}_{I}\right)\,.
			\end{align}
Since these relations involve only operators corresponding to non-scalar
Schur operators in four dimensions, they captures 4d operator relations
that are not visible in the Higgs branch chiral ring.

\subsubsection*{Nulls with $(f,r) = (0,\pm 1)$ and $(\pm 2,0)$}
Similarly, the null operator relations with $f=0$ and $r=\pm 1$ are
			\begin{align}
						D^{+ I}D^{- J}+D^{+J}D^{- I}=0
						\,& , \qquad 
						\bar{D}^{+}_{I}\bar{D}^{-}_{J}+\bar{D}^{+}_{J}\bar{D}^{-}_{I}=0\,, 
						\label{eq:nulld}\\
			\partial X=JX+ \half D^{+I}D^{-}_{I}\,&,\qquad
			\partial \bar{X}= J\bar{X}+\half {\bar{D}^{+}_{I}\bar{D}^{- I}}\,,
			\end{align}
and those with $f=\pm 2$ and $r=0$ are			
			\begin{align}
			\partial^2 B^{\pm}&=2{(T\pm \partial J )B^{\pm}}\mp{J\partial B^{\pm}}
			+{ D^{\pm I}\bar{D}^{\pm}_{I}}\,.\label{eq:difB}
			\end{align}
While \eqref{eq:difB} involves $B^\pm$ arising corresponding to Higgs
branch operators, this operator relation is not captured by 4d Higgs
branch chiral ring relation, since every derivative operator is trivial
in the Higgs branch chiral ring. Therefore, \eqref{eq:difB} captures
more refined data of the OPEs of Higgs branch operators.
		 	
			\subsection{Dim 5}

Let us move on to the null operators at dimension five. Note that there
are such operators that are composite operators involving a null of
dimension four, or the derivatives of lower-dimensional nulls. Below, we
omit all such null operators that trivially follow from the
lower-dimensional ones.

\subsubsection*{Nulls with $f=r=0$}
			
The null operators with $f=r=0$ is 
			\begin{align}
		0 &= 2TC^{A}+(\sigma^{A})^{I}{}_{J}\left(\bar{D}^{+}_{I}\partial
			 D^{-J}+\partial D^{+ J}
			 \bar{D}^{-}_{I}\right)\,,
\\
				0&=J^3T+3J(\partial J)^2-2JT^2	-J\partial( J T)+\half\left(
				 JD^{+ I}\bar{D}^{-}_{I}+\partial D^{- I}\bar{D}^{+}_{I}-\partial D^{+ I}\bar{D}^{-}_{I}
				\right)
\nonumber	\\
				&\qquad +J\partial^2T-\partial J \partial T-\frac{5}{6}J\partial^3 J-4\partial J\partial^{2} J
				+\frac{1}{4}(B^{+}\partial B^{-1}-\partial B^{+1} B^{-1})+\frac{1}{3}\partial^{3} T +\frac{1}{4}\partial^{4}J\,.
			\end{align}
\subsubsection*{Nulls with $f=0$ and $r=\pm 1$}

The null operators of $f=0$ and $r=\pm 1$ are 
			\begin{align}
TX+\frac{1}{4} \left(\partial D^{+I} D^{-}_{I}- D^{+I}\partial D^{-}_{I}\right)=0\,\qquad 
T\bar{X}+\frac{1}{4}\left(\partial\bar{D}^{+}_{I} \bar{D}^{-I}-\bar{D}^{+}_{I} \partial \bar{D}^{-I}\right)=0\,.
			\end{align}

\subsubsection*{Nulls with $f=\pm 1$ and $r=\pm \half $}

The null operator relations of  $f=\pm
1$ and $r=\pm  \half $ are 
			\begin{align}
				2JT D^{\pm I}-\partial^2 JD^{\pm I}-2\partial(J\partial D^{\pm I})+X\bar{D}^{\pm I}
				\pm \left(2J\partial J D^{\pm I}-\partial B^{\pm}D^{\mp I}\right)&=0\,,\\
				2JT \bar{D}^{\pm}_{I}-\partial^2 J\bar{D}^{\pm}_{I}-2\partial(J\partial \bar{D}^{\pm}_{I})-\bar{X}D^{\pm }_{I}
				\pm \left(2J\partial J
			 \bar{D}^{\pm}_{I}-\partial
			 B^{\pm}\bar{D}^{\mp}_{I}\right)&=0\,,\\
			 	C^{A}D^{\pm I}+(\sigma^{A})^{IJ}X\bar{D}^{\pm}_{J} &=0\,,\\
			 C^{A}\bar{D}^{\pm}_{I}-(\sigma^{A})_{IJ}\bar{X}D^{\pm J}& = 0\,.
				\end{align}
			
\subsubsection*{Nulls with $f=\pm 1$ and $r=\pm \frac{3}{2}$}
			
The null operators for $f=\pm 1$ and $r=\pm \frac{3}{2}$ are
			\begin{align}
				XD^{\pm I}=0\,,\quad 
				\bar{X}\bar{D}^{\pm I}=0\,.
			\end{align}

\subsubsection*{Nulls with $f=\pm 2$ and $r=0$}
The null operators with $(f,r) = (\pm 2, 0)$ are 
			\begin{align}
			B^{\pm}C^{A}+	(\sigma^{A})^{J}{}_{I}
			\left(JD^{\pm I}\bar{D}^{\pm}_{J}\mp \partial \left(D^{\pm I}\bar{D}^{\pm}_{J}\right)\right)=0\,.
			\end{align}

\subsubsection*{Nulls with $f=\pm 2$ and $r=\pm 1$}

The null operators for $f=\pm 2$ and $r=\pm 1$ are
			\begin{align}
				B^{\pm}X&= \pm D^{\pm I}\partial D^{\pm}_{I}-\half JD^{\pm I} D^{\pm}_{I}\,,\\
				B^{\pm}\bar{X}&= \pm \bar{D}^{\pm}_{I}\partial \bar{D}^{\pm I}-\half 
				J\bar{D}^{\pm}_{I}\bar{D}^{\pm I}\,.
			\end{align}
			
			\subsection{Dim 6}

We here list independent null operators of dimension six. We again omit
all such operators that trivially follow from lower-dimensional nulls.

\newpage
\subsubsection*{Nulls with $f=r=0$}

There are seven null operators with $f=r=0$:
			\begin{align}
				0=&	T^3-\frac{3}{4}\left(TD^{+ I} \bar{D}^{-}_{I}+\partial D^{+ I} \partial\bar{D}^{-}_{I}+X\bar{X} \right)
					+\half \partial^2 (D^{+ I} \bar{D}^{-}_{I})\nonumber\\
					&-\frac{3}{8}\left(
						D^{+ I} \partial^2\bar{D}^{-}_{I}+\bar{D}^{+}_{I} \partial^2{D}^{-I}
						+JD^{+ I} \partial \bar{D}^{-}_{I}+J\partial \bar{D}^{+}_{I} {D}^{-I}
					\right)\nonumber\\
					&+\frac{1}{4}J^2(\partial^2 T+\partial^3 J)+\frac{3}{2}\left(JT\partial T+\partial J T^2\right)
					+J\partial J \partial T+\frac{5}{4} \left(J \partial^2 JT-\partial J 
					\partial^3J \right) \nonumber\\
				&	+\frac{3}{4}J\partial J \partial^2 J-J\partial^3 T-\frac{9}{4}\partial J \partial^2 T
					-\partial^3 J T -\half \left( J \partial^4 J+(\partial^2 J)^2\right)
					\nonumber\\
					&-\frac{5}{2}T \partial^2 T
			 -(\partial T)^2+\half (\partial J)^2 T
			 -\frac{3}{2}\partial^2 J\partial T
			 +\frac{5}{6}\partial^4 T +\frac{1}{4}\partial^5
			 J\,,
\\[2mm]
0 &= C^AC_A + 6 X\bar{X} + 3\,\partial \left(D^{+I}\partial \bar{D}^-_I +
			 \bar{D}^{+}_I \partial D^{-I}\right) \,,
\\[2mm]
0 &= C^{\{A}C^{B\}} - \frac{1}{3}\delta^{AB}C^DC_D \,, 
			\end{align}
where  $C^{\{A} C^{B\}} :=  \half (C^AC^B +C^BC^A)$.\footnote{For
completeness, we here comment that there are also operator relations of
the form $0 =C^{[A}C^{B]} + i f^{AB C}(\sigma^{C})^{J}{}_{I}\partial
(D^{+I}\partial \bar{D}^{-}_{J}+\bar{D}^{+}_{J}\partial
{D}^{-I})$. These are, however, relations trivially following from the
OPE of $C^A(z)C^B(0)$. We do not list such operator relations as mentioned at
the beginning of this appendix.} Note that the last equation implies five
non-trivial operator relations.

\subsubsection*{Nulls with $f=0$ and $r=\pm 1$}

The null operators with $(f,r) = (0,\pm 1)$ are
			\begin{align}
				XC^{A}=(\sigma^{A})_{IJ}\partial (D^{+I}\partial D^{+J})\,,\qquad
				\bar{X}C^{A}=(\sigma^{A})^{IJ}\partial (\bar{D}^{+I}\partial \bar{D}^{+J})\,.
			\end{align}

\subsubsection*{Nulls with $f=0$ and $r=\pm 2$}

Finally, it turns out that dimension six operators with
$r=\pm 2$ are all null:
			\begin{align}
			X^2=0\,, \quad 
			\bar{X}^2=0\,, 
			\end{align}
			In general, $\DD$ and $\bar{\DD}$ type Schur operators in the Lagrangian theory are
			nilpotent since they involve gauginos \cite{Beem:2013sza,Beem:2017ooy}.
The above relations reflect this property of $\DD$ and $\bar{\DD}$ type
operators in four dimensions.

\bibliography{hoge}
\bibliographystyle{utphys}

\end{document}